\documentclass[aps,10pt,pre,twocolumn,notitlepage,superscriptaddress]{revtex4-1}
\usepackage{graphicx}
\usepackage{float}
\usepackage{amsmath}
\usepackage{mathtools}
\usepackage{color}
\usepackage{xr}
\usepackage{amssymb}
\usepackage{graphicx,color} 
\usepackage{bbm} 
\usepackage{multirow}
\usepackage{enumitem}
\usepackage{mathrsfs}
\usepackage{commath}
\usepackage{hyperref}
\usepackage[dvipsnames]{xcolor}
\hypersetup{colorlinks=true,urlcolor=blue,citecolor=blue,linkcolor=blue}
\newcommand\shortdots{\hbox to 1em{.\hss.\hss.}}

 \let\a=\alpha  \let\g=\psi 
   \let\k=\kappa
 \let\m=\mu   
\let\s=\sigma

\def\de{\mathrm d}

\def\maxphi{\varphi_\mathrm{cp}}
\def\maxphibar{\overline{\varphi}_\mathrm{cp}}
\def\maxrho{\rho_\mathrm{cp}}
\def\minvbar{\overline{v}_\mathrm{cp}}

\def\to{\rightarrow}

\newcommand{\beq}{\begin{equation}} \newcommand{\eeq}{\end{equation}}
\newcommand{\wh}{\widehat}

\begin{document}
\title{Three simple scenarios for high-dimensional sphere packings}
\date{\today}

\author{Patrick Charbonneau}
\affiliation{Department of Chemistry, Duke University, Durham, North Carolina 27708}
\affiliation{Department of Physics, Duke University, Durham, North Carolina 27708}

\author{Peter K. Morse}
\thanks{Corresponding author: peter.k.morse@gmail.com}
\affiliation{Department of Chemistry, Duke University, Durham, North Carolina 27708}

\author{Will Perkins}
\thanks{Corresponding author: math@willperkins.org}
\affiliation{Department of Mathematics, Statistics, and Computer Science, University of Illinois at Chicago, Chicago, Illinois, 60607}

\author{Francesco Zamponi}
\affiliation{Laboratoire de Physique de l'Ecole Normale Sup\'erieure, ENS, Universit\'e PSL, CNRS, Sorbonne Universit\'e, Universit\'e de Paris, F-75005 Paris, France}

\begin{abstract}
Based on results from the physics and mathematics literature which suggest a series of clearly defined conjectures, we formulate three simple scenarios for the fate of hard sphere crystallization in high dimension: (A) crystallization is impeded and the glass phase constitutes the densest packing, (B) crystallization from the liquid is possible, but takes place much beyond the dynamical glass transition and is thus dynamically implausible, or (C) crystallization is possible and takes place before (or just after) dynamical arrest, thus making it plausibly accessible from the liquid state. %Each scenario is associated with a particular bounds on the maximal sphere packing density in high dimensions. 
In order to assess the underlying conjectures and thus obtain insight into which scenario is most likely to be realized, we investigate the densest sphere packings in dimension $d=3$-$10$ using cell-cluster expansions as well as numerical simulations. These resulting estimates of the crystal entropy near close-packing tend to support scenario C.
We additionally confirm that the crystal equation of state is dominated by the free volume expansion and that a meaningful polynomial correction can be formulated.
\end{abstract}

\maketitle

\section{Introduction}

A classical problem of discrete geometry is to determine the maximum fraction of $d$-dimensional Euclidean space, $\maxphi$, that can be covered by non-overlapping, identical spheres.  Determining this densest packing of hard spheres is trivial for $d=1$ and elementary for $d=2$, but %beyond these cases $\maxphi$ is 
otherwise only known rigorously for $d=3$~\cite{hales_proof_2005}, $8$ and $24$~\cite{viazovska_sphere_2017,cohn_sphere_2017}. The behavior of $\maxphi$ as $d \to \infty$ and the structure of the associated packings is a great mathematical challenge about which relatively little is understood.  The best known lower bound is $\maxphi \ge 65963 \cdot d 2^{-d}$~\cite{venkatesh_note_2013} for sufficiently high $d$ (with an additional factor on the order of $\ln (\ln d)$ along a sparse sequence of dimensions), which matches the exponential order of the lower bound $\maxphi \ge 2^{-d}$ trivially obtained by considering any saturated packing~\cite{conway_sphere_1993}.  The best upper bound, by contrast, grows exponentially larger with $d$, as $\maxphi \le 2^{-0.599 d}$~\cite{kabatiansky_bounds_1978,cohn_sphere_2014}. 

Almost all of the known proofs of lower bounds on $\maxphi$ proceed by analyzing lattice packings or random lattice packings (see Ref.~\onlinecite{cohn_packing_2016} for an exposition). These proofs presuppose that lattices provide the backbone of the densest configurations of spheres, but say nothing of the nucleation and coexistence conditions that underlie the ability for a crystal based on such lattices to form and remain stable with respect to the liquid state. While Bravais lattice-based packings are provably optimal in $d = 1$, $2$, $3$, $8$, and $24$, it is far from clear that they remain so for higher $d$~\cite{cohn_packing_2016}. Hence, solely analyzing lattice packings is inadequate to fully capture $\maxphi$.  We here take a statistical physics approach and analyze $\maxphi$ through the equilibrium properties of the hard sphere model, a uniformly random sphere packing of a given density.

We conjecture three possible scenarios for the behavior of the hard sphere model in high dimensions, based on recent work in the physics literature~\cite{radin_structure_2005,koch_most_2005,skoge_packing_2006,van_meel_hard-sphere_2009,estrada_fluidsolid_2011,stevenson_ultimate_2011,charbonneau_thermodynamic_2021,wang_mean-field_2005,finken_freezing_2001,van_meel_geometrical_2009,lue_molecular_2021}:  crystallization (scenario A) does not occur, or, if it does, occurs either (scenario B) much after the dynamical glass transition (at which the liquid dynamics becomes arrested~\cite{parisi_theory_2020}) or (scenario C) around that transition.  Under some simple assumptions and using recent results from both physics and mathematics, we explore the consequences for $\maxphi$ under each scenario.  In A, we conclude that $\maxphi \sim d \ln d \cdot 2^{-d}$; in B, we conclude that $\maxphi $ is only slightly improved to $d^{\g+1} (\ln d)^3 \cdot 2^{-d}$ with some exponent $\g>0$;
in C, we have that $\maxphi \ge 2^{-d(1-\epsilon)}$ for some explicit $\epsilon >0$. It is worth noting that Refs.~\onlinecite{torquato_new_2006, torquato_jammed_2010} proposed a series of conjectures based on a different set of arguments from statistical physics, that are consistent with scenario C. See also Refs.~\onlinecite{kallus_statistical_2013,andreanov_extreme_2016} and references therein. 

A set of plausibility conditions emerge for each of these scenarios, which are then checked against simulation results and a cell-cluster expansion of the densest known crystals in $d=3$-$10$, which are expected to be the most thermodynamically stable at high pressures (and have been observed to be so for all pressures at which crystals are stable in $d=3$-6~\cite{van_meel_hard-sphere_2009, charbonneau_thermodynamic_2021, lue_molecular_2021}). The inclusion of $d=10$ here is significant, as it is the lowest dimension for which a non-Bravais lattice is the basis for the densest known crystal. While these observations do not suffice to unambiguously declare which scenario is correct, they nevertheless suggest that scenario C is most likely, followed by scenario B. While scenario A remains plausible, no hint of it can be teased from low-dimensional crystallization trends. 

The rest of this article is organized as follows. In Section~\ref{sec:background}, we provide a series of definitions and describe the first-order liquid-crystal phase transition in hard spheres. In Section~\ref{sec:conjectureSection}, we present the aforementioned conjectures and follow through with their implications for three possible crystallization scenarios. 
In Section~\ref{sec:simulations}, we analyze low-dimensional crystals in $d=3$-$10$ using cell cluster expansions (where numerically possible) as well as simulations to further constrain the likely scenarios. Section~\ref{sec:final} concludes with a discussion of the likelihood of each of the three scenarios given the low-dimensional trends.

\section{Theoretical Background}
In this section, we provide a definition of the entropy of the hard sphere model and show that both its first and second derivatives with respect to volume are positive. We then use these properties to derive the relationship between liquid and crystal entropies through a common tangent construction.
\label{sec:background}
\subsection{Definitions}

Consider $N$ identical $d$-dimensional hard spheres of diameter $\sigma$ in a box of volume $V$.
Sphere positions are specified by a set of $d$-dimensional vectors $\underline{Y} = \{ \mathbf{y}_i \}_{i=1,\cdots,N}$, each $\mathbf{y}_i$ having components $y_{i\m}$ for $\m=1,\cdots,d$.
The sphere concentration is equivalently described by the number density $\rho = N/V$, the specific volume $v = 1/\rho = V/N$, and the packing fraction $\varphi = \rho V_d (\sigma/ 2)^d$, where
$V_d = \pi^{d/2}/\Gamma(1+d/2)$ is the $d$-dimensional volume of a ball of unit radius. In the following, we consider the thermodynamic limit in which $N\to\infty$ and $V\to\infty$, at constant $\varphi \in (0, \maxphi)$.

Defining $I(\underline{Y})$ the indicator function specifying that there are no overlaps between spheres, one can introduce 
\begin{align}
Z_N &= \frac1{N!} \int \de \underline{Y}  \, I(\underline{Y}) \ , \qquad Z^{\rm id}_N = \frac{V^N}{N!} \ , \\
Z^{\rm ex}_N &= \frac1{V^N} \int \de \underline{Y} \, I(\underline{Y})  = \frac{Z_N}{Z^{\rm id}_N} \nonumber \ ,
\end{align}
which are the configurational, ideal gas, and excess partition functions, respectively. Note that $Z^{\rm ex}_N \in [0,1]$
is also the probability that $N$ randomly placed spheres in $V$ have no overlap.
Similarly, in the thermodynamic limit, the per particle, ideal gas, and excess entropies are, respectively,
\begin{align}
s &=\lim_{N\to\infty} \frac1N \ln Z_N = s^{\rm id}+ s^{\rm ex} \ , \nonumber \\ \label{eq:excessS}
s^{\rm id} &= -\ln(\rho\sigma^d) - d\ln (\Lambda/\sigma) + 1 \ ,  \\ 
s^{\rm ex} &=\lim_{N\to\infty} \frac1N \ln Z^{\rm ex}_N \ , \nonumber
\end{align}
where $\Lambda$ is the de Broglie wavelength and the sphere diameter $\sigma$ is here introduced purely for notational convenience.
The thermodynamic relations for pressure $P$ and isothermal compressibility $\chi_T$ (for temperature $T=1/\beta$ with the Boltzmann constant set to unity)
\begin{equation}\label{eq:thermo}
\beta P = \frac{ds}{dv} \geq 0 \ ,
\qquad
\chi_T = - \frac1V \frac{dV}{dP} = -\frac{\rho}{T} \frac{1}{\frac{d^2s}{dv^2}} \geq 0 \ ,
\end{equation}
imply that the total entropy per particle, $s$, must be a monotonically increasing and concave function of the specific volume.

\subsection{Crystallization via a first-order phase transition}

\begin{figure}[t]
\includegraphics[width=\columnwidth]{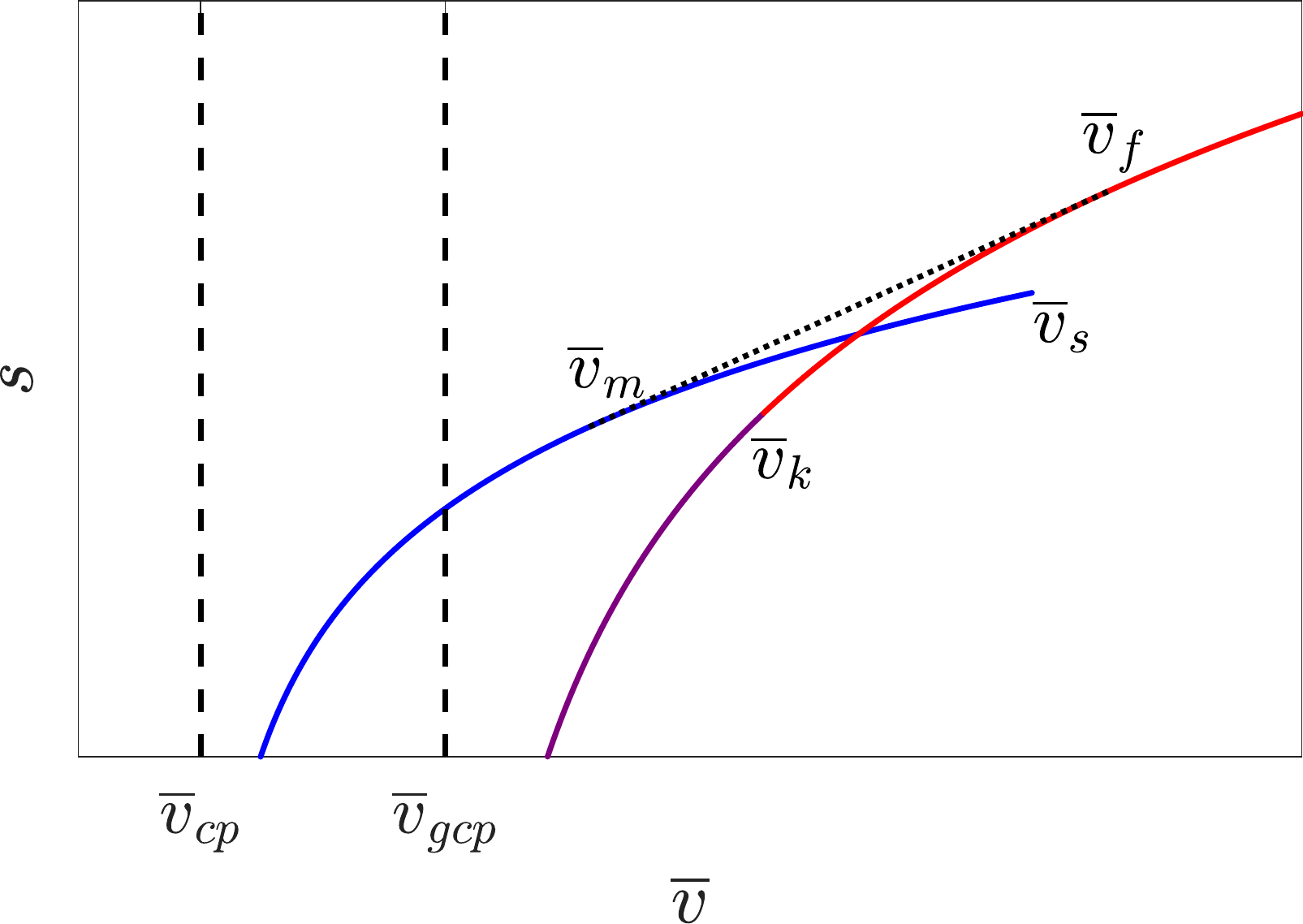}
\caption{Sketch of the liquid (red) and crystal (blue) entropies as a function of the scaled specific volume $\overline{v}$, in the vicinity of the first-order fluid-crystal transition determined by common tangent construction (black dotted line). The crystal branch terminates at the densest packing density ${\maxphibar=1/\minvbar}$ (dashed line) and remains metastable beyond coexistence up to $\overline{\varphi}_s = 1/\overline{v}_s$. % the minimum metastable density for a crystal. 
For $d=3$-$10$, $\overline{\varphi}_s>\overline{\varphi}_f$~\cite{charbonneau_thermodynamic_2021}, but no assumption is here made about their ordering in higher $d$. The liquid branch extends from zero density, and its metastable extension beyond coexistence terminates at the Kauzmann density ${\overline{\varphi}_{k}=1/\overline{v}_k}$, whereupon the liquid turns into an ideal glass (purple). The glass phase then terminates at the glass close packing density ${\overline{\varphi}_{\rm gcp}=1/\overline{v}_{\rm gcp}}$~\cite{parisi_theory_2020}.}
\label{fig:sketch}
\end{figure}

In all dimensions $d$ for which the information is available, the densest (infinite-pressure) packing of hard spheres is crystalline; that is, given by a (Bravais or not) lattice packing. For $d\geq3$, at finite pressure, this densest packing gives rise to a stable crystalline phase separated from the liquid phase by a first-order transition. Such a liquid-crystal transition means that the liquid and crystal phases have distinct analytic entropy functions,
$s_{\ell}$ and $s_{c}$, which are separately monotonically increasing and concave.
Because Eqs.~\eqref{eq:thermo} should always be satisfied in equilibrium, the equilibrium state of the system corresponds to the Maxwell construction illustrated in Fig.~\ref{fig:sketch}.
At low $P$ (high $v > v_f$), the homogeneous liquid dominates; at high $P$ (low $v < v_m$),
the homogeneous crystal dominates. In the region $v_m < v < v_f$, pressure $P_\mathrm{co}$ is constant and the system is formed of coexisting crystalline and liquid domains.
The equations determining the three unknown $v_m, v_f, P_\mathrm{co}$ which characterize the
coexistence region can be obtained from the common tangent construction defined as
\begin{equation}
\begin{split}
&\frac{ds_\ell}{dv}(v_f) = \frac{ds_c}{dv}(v_m) = \beta P_\mathrm{co} \ , \\
&s_\ell(v_f) - s_c(v_m) = \beta P_\mathrm{co} (v_f - v_m) \ .  
\end{split}
\label{eq:coex}
\end{equation}

\section{Conjectures and Scenarios}
\label{sec:conjectureSection}
In this section, we describe a set of conjectures that constrain the relationships in Eq.~\eqref{eq:coex} and work through their consequences, hence giving rise to three crystallization scenarios. Note that in considering high-$d$ systems, it is convenient to define
scaled packing fraction $\overline{\varphi} =2^d \varphi = \rho V_d \sigma^d$,
specific volume $\overline{v} = 1/\overline{\varphi} = v/(V_d \s^d)$,
and pressure $\overline{P} = \beta P \sigma^d / V_d$. 

\subsection{Conjectures on the high-\texorpdfstring{$d$}{Lg} phase behavior}
\label{sec:conj}
 %We define the following threshold fractions  $\overline{\varphi}_{l}\le \overline{\varphi}_{k}\le \overline{\varphi}_{c} = \overline{\varphi}_{\rm gcp} \le  \maxphibar = 2^d \maxphi$ below.
We first make a series of conjectures:
\begin{enumerate}

\item The high $d$ equilibrium phase diagram is characterized by a low-density liquid and a high-density crystal, and no other equilibrium phase intervenes.  If there is a high-density crystal phase, then it is separated from the liquid phase by a first-order phase transition, as in Fig.~\ref{fig:sketch}.  We have no support for this conjecture, other than from the empirical observation that it holds in $d=3$-$10$~\cite{skoge_packing_2006,van_meel_hard-sphere_2009, charbonneau_thermodynamic_2021, lue_molecular_2021}.

\item In the limit of high $d$, the excess entropy of the liquid phase is given by truncating the virial expansion at the lowest order, i.e.,
\begin{equation}
s^{\rm ex}_\ell = - \frac{\overline{\varphi}}{2} = -\frac{1}{2 \overline{v}}  \ ,
\end{equation}
which implies that
\begin{equation}
\label{eq:Sliq}
s_\ell = \ln\overline{v} -\frac{1}{2 \overline{v}} + \ln V_d - d \ln (\Lambda/\sigma) + 1 \ .
\end{equation}

Equation~\eqref{eq:Sliq} holds up to the so-called {\it Kauzmann density} $\overline{\varphi}_k = d \ln d + o(d\ln d)$, at which the liquid state condenses into an ideal glass
phase. The ideal glass entropy has a different and less explicit expression---see Ref.~\onlinecite[Eq.~(7.43)]{parisi_theory_2020} and the surrounding discussion---that is continuous at $\overline{\varphi}_k$ and quickly diverges to $-\infty$ upon approaching the {\it glass close packing} density
${\overline{\varphi}_{\rm gcp} = d \ln d + o(d\ln d)}$, as illustrated in Fig.~\ref{fig:sketch}.
(The difference between $\overline{\varphi}_k$ and
$\overline{\varphi}_{\rm gcp}$ is at the level of subleading corrections.)
In addition, the liquid dynamics become arrested for ${\overline{\varphi} > \overline{\varphi}_d \approx 4.8 d}$.
This conjecture is supported by a large body of physics literature~\cite{frisch_classical_1985, wyler_hard-sphere_1987, frisch_high_1999, parisi_toy_2000, parisi_mean-field_2010,
maimbourg_solution_2016, charbonneau_glass_2017, parisi_theory_2020, charbonneau_dimensional_2021}.

\item The crystal phase is accurately described by the free-volume entropy. In other words, throughout the crystal phase, particles simply
rattle in a cage formed by their neighbors. Consider the close packed crystal at density $\maxphibar$, and reduce the diameter of all particles from $\sigma$ to
$\sigma(1-\varepsilon)$. The density is correspondingly reduced to $\overline{\varphi} = \maxphibar (1-\varepsilon)^d$, and each particle gains the possibility of rattling in a volume of linear size
$a \varepsilon\sigma$ without overlapping its neighbors, $a$ being an unknown proportionality constant close to 1.
Moreover all particles can be permuted, so each particle can access all the $N$ possible cages.
Therefore, using $x=\minvbar/\overline{v}=\rho/\maxrho$, one can estimate
\begin{align}
\label{eq:cryst}
Z^{\rm ex}_N \approx& \left[ \frac{N V_d  (a \varepsilon\sigma)^d  }{V} \right]^N = \left[ \maxphibar (a \varepsilon)^d \right]^N
\Rightarrow\nonumber\\
s_c \approx& -\ln x + d \ln a + d \ln ( 1 - x^{1/d}) \\ &+\ln V_d - d \ln (\Lambda/\sigma) + 1 \ . \nonumber
\end{align}
We assume that this expression remains valid for all $\overline{v} \in [ \minvbar, \overline{v}_m ]$.
We have no support for this conjecture, except from the empirical observation that a similar expression provides a good fit to the crystal entropy in $d=3$-$10$~\cite{van_meel_hard-sphere_2009, charbonneau_thermodynamic_2021, lue_molecular_2021}.  The free volume entropy gives a rigorous lower bound on $s^{\rm ex}$, and, if we assume the close-packed crystal to be a lattice packing, then we can allow particles to rattle in regions defined by scaling down the Voronoi cells around each center. A special consideration should be made for lattice packings, such as $\lambda_9$, which contain a set of internal soft (or zero) modes. Along such modes, the packing is allowed to shift freely without generating any overlap. Because the number of such modes is necessarily subextensive, however, the contribution of these modes to the entropy per particle must vanish in the thermodynamic limit (by analogy to the contribution of Goldstone modes in the low-temperature phase of a Heisenberg ferromagnet \cite{patashinskii_fluctuation_1979}).

\item The liquid remains the equilibrium phase at least down to a specific volume $\overline{v} = 1/[d \ln(2/\sqrt{3})]$, i.e., $\overline{v}_\ell < 1/[d \ln(2/\sqrt{3})] \approx 6.952/d$
or $\overline{\varphi}_\ell > d \ln(2/\sqrt{3}) \approx 0.144 d$.
This conjecture is motivated by the results of~\cite{jenssen_hard_2019}.
\end{enumerate}

\subsection{Crystallization in high \texorpdfstring{$d$}{Lg}}

From the conjectures of Sec.~\ref{sec:conj} we can derive bounds on high-$d$ crystallization, which are discussed below.

\subsubsection{Coexistence equations}

First, by rewriting Eqs.~\eqref{eq:coex} in terms of scaled variables and using Eq.~\eqref{eq:Sliq} for $s_\ell$ and Eq.~\eqref{eq:cryst} for $s_c$, we obtain
\begin{equation}
\begin{split}
&\frac{1}{\overline{v}_f} + \frac{1}{2 \overline{v}_f^2} =
 \frac1{\overline{v}_m} \frac{1}{ 1 - \left( \minvbar/\overline{v}_m \right)^{1/d} }
= \overline{P}_\mathrm{co} \ , \\
&\ln(\overline{v}_f) - \frac{1}{2 \overline{v}_f}  - \ln( \overline{v}_m /  \minvbar ) - d \ln a \\
& - d \ln \left[ 1 - \left( \frac{\minvbar}{\overline{v}_m} \right)^{1/d} \right]  = \overline{P}_\mathrm{co} (\overline{v}_f - \overline{v}_m) \ .  
\end{split}
\end{equation}
It is then convenient to rewrite these equations in terms of density $\overline{\varphi}$:
\begin{equation}\label{eq:coexolf}
\begin{split}
&\overline{\varphi}_f + \frac{1}{2} \overline{\varphi}_f^2 =
  \frac{\overline{\varphi}_m}{ 1 - \left( \overline{\varphi}_{m}/\maxphibar \right)^{1/d} }
= \overline{P}_\mathrm{co} \ , \\
&-\ln(\overline{\varphi}_f) - \frac{1}{2} \overline{\varphi}_f  - \ln( \maxphibar /  \overline{\varphi}_{m} ) - d \ln a \\
& - d \ln \left[ 1 - \left( \frac{\overline{\varphi}_{m}}{\maxphibar} \right)^{1/d} \right]  =
 \overline{P}_\mathrm{co} \left( \frac{1}{\overline{\varphi}_f} -\frac{1}{\overline{\varphi}_m} \right) \ .  
\end{split}
\end{equation}
Given $\maxphibar$, these equations can easily be solved numerically to yield the coexistence parameters. This strategy was employed by Finken et al.~\cite{finken_freezing_2001} (albeit possibly with an erroneous common tangent construction~\cite{van_meel_hard-sphere_2009}) using close packing density of laminated lattices up to $d\approx 50$. Here we take a different approach. We use our knowledge of $\overline{\varphi}_f$ to obtain bounds on $\maxphibar$.

\subsubsection{Asymptotic analysis}
According to Sec.~\ref{sec:conj}, one has $\overline{\varphi}_f \in [0.144 d,  d \ln d]$. In a more strict setting we could impose that crystallization happens
before the liquid is dynamically arrested, which would restrict the upper bound to $4.8 d$.  
We thus introduce $\widehat{\varphi}_f = \overline{\varphi}_f/d$ that is of $\mathcal{O}(1) $ or at most $\mathcal{O}(\ln d)$.
For $d\to\infty$, we have $\ln(\overline{\varphi}_f) \ll \frac{1}{2} \overline{\varphi}_f$ and $\overline{\varphi}_f \ll \frac{1}{2} \overline{\varphi}_f^2$, and also $\overline{P}_\mathrm{co} \sim \frac{1}{2} \overline{\varphi}_f^2$,
which thus simplifies Eqs.~\eqref{eq:coexolf} as:
\begin{equation}
\begin{split}
& \frac{d^2}2 \widehat{\varphi}_f^2 =
 \frac{\overline{\varphi}_m}{ 1 - \left( \overline{\varphi}_{m}/\maxphibar \right)^{1/d} }  \ , \\
& - d \widehat{\varphi}_f  - \ln( \maxphibar /  \overline{\varphi}_{m} ) \\
& - d \ln a - d \ln \left[ 1 - \left( \frac{\overline{\varphi}_{m}}{\maxphibar} \right)^{1/d} \right]  =
- \frac{d^2}{2} \frac{ \widehat{\varphi}_f^2}{ \overline{\varphi}_m} \ .  
\end{split}
\label{eqn:asymptotics}
\end{equation}
Two possible asymptotic solutions to these equations exist, depending on the scaling of $\overline{\varphi}_m / \maxphibar$.

\subsection{Crystallization scenarios}
\label{sec:3scenarios}
Based on the above conjectures and asymptotic analysis, three distinct crystallization scenarios can be identified.

\subsubsection{Scenario A}

In this scenario, crystallization does not proceed and thus the liquid and the glass phases are the only possible equilibrium phases. The close packing density then equals the glass close packing density, and hence $\maxphi = 2^{-d} \cdot \overline{\varphi}_{\rm gcp} \sim 2^{-d} d \ln d$. This scenario happens if the close packing density of the densest crystal remains below $\varphi_{\rm gcp}$.

\subsubsection{Scenario B}

In this scenario, we suppose that there is a crystalline phase and  ${\overline{\varphi}_m / \maxphibar \sim A/d^\g}$ (with $\g>0$ and $A>0$, or $\g=0$ and $0<A<1$), such that ${1-\big(\overline{\varphi}_m/\overline{\varphi}_\mathrm{cp}\big)^{1/d} 
%=1 - \left( \overline{\varphi}_{m}/\maxphibar \right)^{1/d} 
\sim ( \g \ln d - \ln A)/d}$.
Note that in this scenario $1-\big(\overline{\varphi}_m/\overline{\varphi}_\mathrm{cp}\big)^{1/d} \allowbreak \ll 1$ and the use of the free volume equation of state for the crystal is well justified.
Defining $\widehat{\varphi}_m  = \overline{\varphi}_m/d$
and neglecting subdominant terms, Eqs.~\eqref{eqn:asymptotics} become
\begin{equation}
\begin{split}
& \frac{1}{2} \widehat{\varphi}_f^2 =
 \frac{\widehat{\varphi}_m }{  \g \ln d - \ln A} \ , \\
& -  \widehat{\varphi}_f   -  \ln a - \ln(\g \ln d - \ln A) + \ln d   =
- \frac{1}{2} \frac{ \widehat{\varphi}_f^2}{ \widehat{\varphi}_m}  \ .  
\end{split}
\end{equation}
The solution is
\begin{equation}
\label{eq:Bris}
\begin{split}
\overline{\varphi}_f &\sim d \ln d \ , \\
\overline{\varphi}_m &\sim d \frac{(\ln d)^2}{2} ( \g \ln d - \ln A) \ , \\
\maxphibar &\sim \frac{d^{\g+1}}{A}  \frac{(\ln d)^2}{2} ( \g \ln d - \ln A) \ . \\
\end{split}
\end{equation}
Note that one should check the subleading corrections to $\overline{\varphi}_f$ to make sure that $\overline{\varphi}_f \leq \overline{\varphi}_k$, which is a strict requirement for the consistency of our approach.
%\item 
It is also somewhat unpleasant that crystallization then takes place much beyond the dynamical arrest of the liquid, i.e., $\overline{\varphi}_f \gg \overline{\varphi}_d$. %becomes dynamically arrested.
%\end{itemize}
In this scenario, the close-packed crystal would be only slightly denser than the best amorphous packing, and its exponential scaling would
be the same as the Minkowski bound. Crystallization would then be extremely unlikely, because the liquid would becomes dynamically arrested before any sign of crystallization could emerge. Note that the value of $a$, provided it remains finite for $d\to\infty$, here plays no role.

\subsubsection{Scenario C}

In this scenario, we suppose there is a crystalline phase, but by contrast to scenario B, here $\overline{\varphi}_m / \maxphibar \sim e^{-\a d}$ with constant $\alpha > 0$.
Hence, $1-\big(\overline{\varphi}_m/\overline{\varphi}_\mathrm{cp}\big)^{1/d} = 1-e^{-\a}$ remains finite, and the use of the free volume equation of state for the crystal is less justified for large $\a$.
Then, the first equation gives $\overline{\varphi}_m = (d^2/2) \widehat{\varphi}_f^2 (1-e^{-\a})$.
Plugging this expression into the second Eq.~\eqref{eqn:asymptotics} and taking the leading order, we get
$\widehat{\varphi}_f = -\a -\ln(1-e^{-\a}) - \ln a$.
The final result is then
\begin{equation}
\label{eq:finC}
\begin{split}
& \overline{\varphi}_f \sim d [ -\a -\ln(1-e^{-\a}) - \ln a ] \ , \\
& \overline{\varphi}_m \sim \frac{1}{2} \overline{\varphi}_f^2 (1-e^{-\a}) \ , \\
&\maxphibar \sim e^{\a d}  \overline{\varphi}_m \ . \\
\end{split}
\end{equation}
In this scenario, the beginning of the coexistence region is $\overline{\varphi}_f \propto d$, which is a natural scaling for the liquid state, while the end of the coexistence region
is $\overline{\varphi}_m \propto d^2$ and the crystal close packing is $\maxphibar \propto e^{\a d}$.

\begin{figure}[t]
\includegraphics[width=\columnwidth]{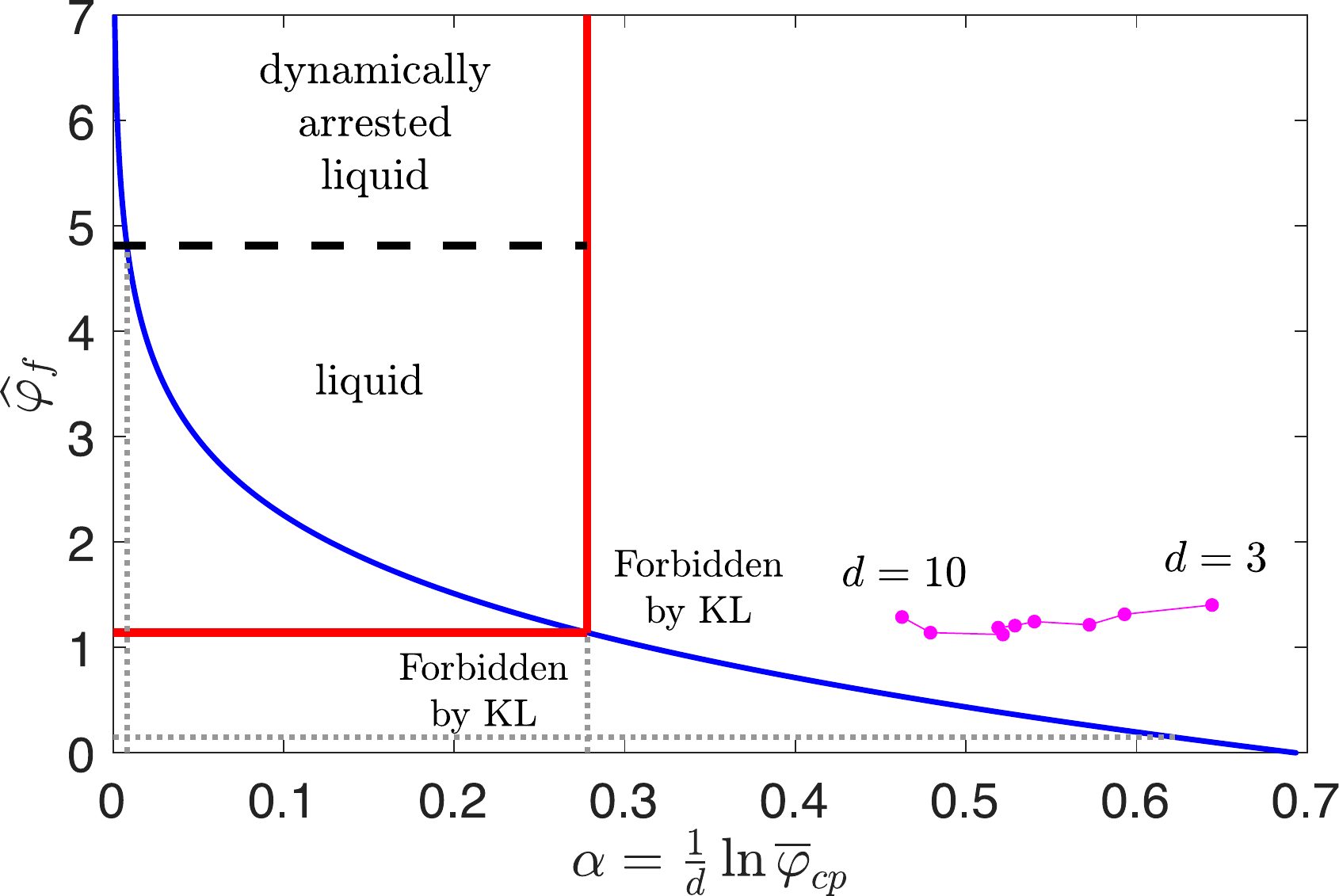}
\caption{Sketch of $\widehat{\varphi}_f$ as a function of $\a=\frac{1}{d}\ln\maxphibar$ in scenario C, assuming that $\ln a=0$. All areas to the right and below of the red box are forbidden by the KL bound when ${d\to\infty}$. Simulation results for the freezing density in $d=3$-$10$ (magenta) trend towards the asymptotically allowed region as $d$ increases.}
\label{fig:bound}
\end{figure}

The relation between $\widehat{\varphi}_f =\overline{\varphi}_f/d$ and $\alpha=\frac{1}{d}\ln\overline{\varphi}_\mathrm{cp}$
(at fixed $a$), given
in Eq.~\eqref{eq:finC}, 
is illustrated in Fig.~\ref{fig:bound} (for $a=1$).
It is a decreasing function, and it
can be inverted to
give
\begin{equation}
    \alpha = \ln(1+e^{-\widehat{\varphi}_f-\ln a}) \ .
\end{equation}
Hence, upper bounds on $\alpha$ can be turned into lower bounds for $\widehat{\varphi}_f$, and vice versa.

Let us consider upper bounds on $\a$ first. The Kabatiansky-Levenshtein (KL) upper bound on packing~\cite{kabatiansky_bounds_1978} requires that $\a \leq \allowbreak \ln(2) (1-0.5990) \allowbreak = 0.278$, which then implies
\begin{equation}\label{eq:wfflb}
\widehat{\varphi}_f > 1.138 -\ln a \ .
\end{equation}
The fourth
conjecture in Sec.~\ref{sec:conj}
implies a bound $\widehat{\varphi}_f \geq 0.144$;
as long as $\ln a \leq 0.994$, this bound
is however weaker than the KL one.

We then consider lower bounds on $\a$. 
If we require that crystallization happens before dynamical arrest, 
i.e. $\widehat{\varphi}_f\leq 4.8$, then
we obtain a lower
bound on $\alpha$ in terms of~$a$,
\begin{equation}
\ln\big(1+e^{-4.8-\ln a}\big) \le \alpha \ .
\label{eqn:alphaRange}
\end{equation}
Note that crystallization might well take place after
dynamical arrest, as it is the case in scenario B discussed above; if $\ln a < -3.662$, then this is necessarily the case, due to Eq.~\eqref{eq:wfflb}.

Unfortunately, $a$ is unknown, but if the third conjecture of Sec.~\ref{sec:conj} is correct, then it should be close to unity, which is what our finite-$d$ results suggest (see Sec.~\ref{sec:simulations}). Because the dependence of the above bounds on $a$ is logarithmic, relatively small deviations from unity further do not affect much the results.
We thus assumed $a \rightarrow 1$ as 
$d \rightarrow \infty$, for illustration, 
in Fig.~\ref{fig:bound}.
With this choice, we obtain 
$\wh{\varphi}_f = -\a -\ln(1-e^{-\a})$ (blue line)
and
\begin{equation}
\ln(1+e^{-4.8}) =0.0082 \leq \a \leq 0.278 \ .
\end{equation}
The true value of $a$ in the high-dimensional limit simply sets the ordinate offset of the blue curve
in Fig.~\ref{fig:bound} and, provided it is not too far from unity, only slightly shifts the lower bound.

\subsubsection{Summary of the three scenarios}

From the analysis so far, we conclude that under the assumptions in Sec.~\ref{sec:conj}, three possible scenarios arise:
\begin{itemize}
\item[A.] If there is no crystallization, then $\maxphi = 2^{-d} \overline{\varphi}_{\rm gcp} \sim d \ln d \cdot 2^{-d}$, and optimal packings are glasses.
\item[B.] If the close packing density of the crystal is not exponentially larger than the melting density, then crystallization happens deep in the dynamically arrested region ($\overline{\varphi}_d \ll \overline{\varphi}_f < \overline{\varphi}_k \sim d \ln d$), and we obtain the results in Eq.~\eqref{eq:Bris},
in particular with $\maxphibar \sim d^{\g+1}  \frac{(\ln d)^2}{2} ( \g \ln d - \ln A)$ being not exponential in $d$, and only slightly larger than $\overline{\varphi}_{\rm gcp}$.
\item[C.] If instead crystallization happens on the same scale as the dynamical arrest ($\overline{\varphi}_f\propto d$),
then the crystal close packing should be $\maxphibar\sim e^{\a d}$.
Quantitative bounds then depend weakly (logarithmically) on $a$; we assume $a=1$ for simplicity, 
which gives an upper bound $\a < 0.278$ from the KL bound and also implies $\overline{\varphi}_f > 1.138 d$.
The additional requirement that crystallization happens {\it before} dynamical arrest
($\overline{\varphi}_f < \overline{\varphi}_d\sim 4.8d$) gives a lower bound $\a>0.0082$ on the close packing density (assuming $\ln a \approx 0$), which improves exponentially over the Minkowski bound.
\end{itemize}

\section{Insights from low-\texorpdfstring{$d$}{Lg} crystals}
\label{sec:simulations}

To obtain insights into which of the above three scenarios is most likely, we examine $\overline{\varphi}_f$, $\overline{\varphi}_m$, and $a$ for each of the densest crystals in $d=3$-$10$, which are $D_d$ checkerboard lattices in $d=3$-$5$, $E_6$ in $d=6$, $E_7$ in $d=7$, the $E_8$ root lattice in $d=8$, $\lambda_9$ in $d=9$, and $P_{10c}$ in $d=10$~\cite{conway_sphere_1993}. Because the relative distance between $\overline{\varphi}_m$ and $\overline{\varphi}_\mathrm{cp}$ grows with $d$, it is possible (through corrections to the free-volume expressions considered here) that a lower density crystal may be most stable at intermediate pressures for certain $d$.  We here consider only the densest crystals, as all other crystal forms would in any case transition to the densest at high pressure given enough time. (Note that low-dimensional studies in $d=2$-$6$ have found no such discrepancy~\cite{van_meel_hard-sphere_2009, lue_molecular_2021}, and that if one were to exist, the densest crystal would nevertheless offer the strongest bound on the stability of the liquid, and thus the scaling analysis would not be impacted.)

In this section, we consider three distinct estimates of the constant $a$ for these crystals: (1) a high-$d$ generalization of the Rudd--Stillinger cell-cluster expansion~\cite{rudd_rigid_1968} for nearly perfect crystals; (2) the scaling of the dynamical cage size near close packing; (3) thermal integration of the crystal equation of state from a reference crystal whose absolute entropy is determined by the Frenkel-Ladd scheme~\cite{frenkel_new_1984}.

\begin{table*}[htb]\centering
\begin{tabular}{c | c |c | c | c | c | c | c}
\hline
\hline
crystal & $\maxphi$ & $\kappa_0^\mathrm{CC_1}$ & $\kappa_0^\mathrm{CC_2}$ (Dir) & $\kappa_0^\mathrm{fit}$ & $\kappa_1^\mathrm{CC_1}$ & $\kappa_1^\mathrm{CC_2}$ (Dir) & $\kappa_1^\mathrm{fit}$\\
\hline
$D_3$ & 0.7405 & 0.125 & 0.3136 & 0.511(18) & 0.6115 & 2.2108 & 3.8(4)\\
$D_4$ & 0.6169 & 0.15 & 0.3403 & 0.491(18) & 0.76 & 2.4845 & 3.3(3)\\
$D_5$ & 0.4653 & 0.2 & 0.4299 & 0.555(12)& 0.9205 & 1.5717 & 3.2(2)\\
$E_6$ & 0.3729 & 0.4286 & -- & 0.54(2) & 0.1228 & -- & 2.8(3)\\
$E_7$ & 0.2953 & 0.4554 & -- & 0.48(3) & -- & -- & 3.2(4)\\
$E_8$ & 0.2537 & 0.4336 & -- & 0.41(3)& -- & -- & 3.1(5)\\
$\lambda_9$ & 0.1458 & 0.4370 & -- & 1.2(2) & -- & -- & -3(2)\\
$P_{10c}$ & 0.0996 & -- & -- & 0.62(5) & -- & -- & 3.1(7)\\
\hline
\hline
\end{tabular}
\caption{\textbf{Constants used and derived for each crystal.} Packing fraction at close packing $\maxphi$ taken from Ref.~\onlinecite{conway_sphere_1993}. Cell cluster equation of state results are given to both first ($\kappa_0^\mathrm{CC_1}$ and $\kappa_1^\mathrm{CC_1}$) and second ($\kappa_0^\mathrm{CC_2}$ and $\kappa_1^\mathrm{CC_2}$) order, and are compared with the crystal equation of state obtained from numerical simulations, $\kappa_0^\mathrm{fit}$ and $\kappa_1^\mathrm{fit}$. Cell cluster results are exact to machine precision, but rounded to the fourth decimal places. Otherwise, error bars represent $95\%$ confidence intervals.}
\label{table:packingConstants}
\end{table*}

\subsection{Cell-cluster expansion}
\label{sec:cellClusterMain}
Rudd and Stillinger proposed to expand the entropy of a high-pressure crystal of hard particles~\cite{rudd_rigid_1968} by ordering terms as
\begin{multline}
s_c = \lim_{x\to 1} \bigg[-d\ln(\Lambda/\sigma) + d\ln(1 - x^{1/d}) - \ln x - C \\ - D(1-x^{1/d})
- E(1-x^{1/d})^2 + \mathcal{O}(1-x^{1/d})^3 \bigg] \ ,
\label{eqn:rudd}
\end{multline}
where $x=\varphi/\maxphi$. In essence, this scheme proposes a polynomial correction in $1-x^{1/d}$ to the free volume expansion of Eq.~\eqref{eq:cryst}. The coefficients $C$, $D$, and $E$, which depend on crystal symmetry and dimension, can further be expanded in (infinite) series of cell clusters (see Appendix~\ref{sec:cellClusterAppendix}). We here present two such expansions, denoted recursive (Rec) and direct (Dir). Although these series are neither unique nor proven to converge in any $d$, they nevertheless provide a constructive analytical framework. By (admittedly loose) physical analogy with the virial expansion for the liquid, one might even expect their convergence rate to improve as $d\rightarrow\infty$. 

Using Eqs.~\eqref{eq:thermo} and~\eqref{eqn:rudd}, the reduced pressure (or compressibility) can then be expressed as
\begin{align}
p &= \frac{\beta P}{\rho} =-x\left(\frac{\partial s_c}{\partial x}\right)_{\beta}\nonumber\\
%&= \frac{\beta P}{\rho} = \frac{1}{\rho}\left(\frac{\partial s}{\partial 1/\rho}\right)_{\beta} \nonumber 
 &= \frac{1}{1-x^{1/d}} + \kappa_0 + \kappa_1(1-x^{1/d}) + \mathcal{O}(1-x^{1/d})^2,
\label{eqn:compressDeriv}
\end{align}
where the first term $1/(1-x^{1/d})$ is the free volume equation of state~\cite{kirkwood_critique_1950, kamien_entropic_2007}, and its constant and linear offsets corrections are, respectively, 
\begin{align}
\kappa_0 &= - \frac{D}{d} \ ,\label{eq:relationK0} \\
\kappa_1 &= \frac{D-2E}{d} \ .\label{eq:relationK1}
\end{align}
Comparing Eqs.~\eqref{eq:cryst} and \eqref{eqn:rudd} further identifies
\begin{equation}
d \ln a = - C - 1 -\ln V_d \ .
\label{eq:relationCA}
\end{equation}
Values of $\kappa_0$ and $\kappa_1$ from the cell cluster expansions are presented in Table~\ref{table:packingConstants}.

\begin{figure}[htb]
\includegraphics[width=0.5\linewidth]{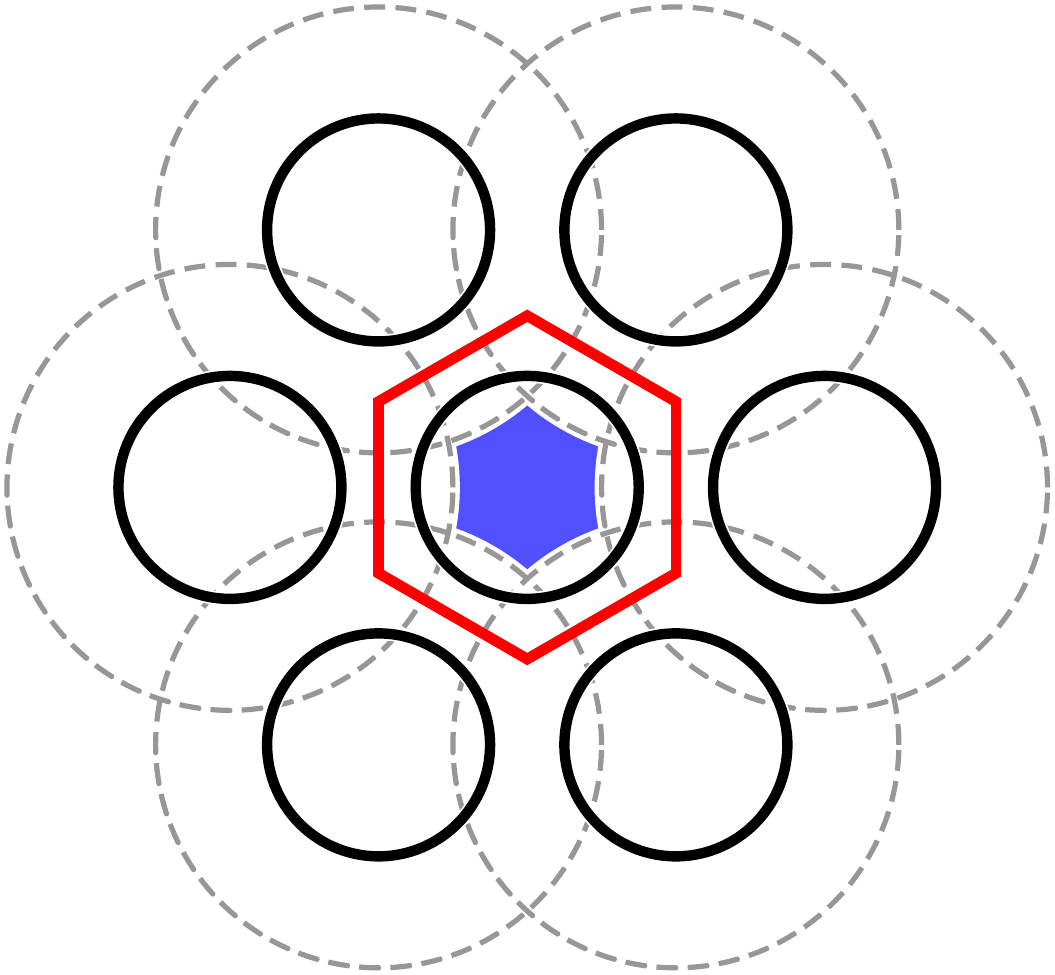}
\caption{Free volume schematic with each sphere (black circle) excluding a spherical volume of radius $\sigma$ (dashed lines) away from their centers for the center of another sphere to occupy. The space thus bounded (blue dashed lines) is the free volume, $v_\mathrm{free}$ (blue), available to the center sphere. For $x\to1$, the free volume boundary is approximately self-similar to that of the Voronoi cell (red), but upon decompression the curvature of the free volume boundary grows more pronounced. This concavity implies $\frac{\partial^2 v_\mathrm{free}}{\partial x^2}<0$, and thus $\kappa_0 > 0$.}
\label{fig:curvedFV}
\end{figure}

The standard derivation of the free volume---and $s_c$ by extension---assumes that upon decompressing a close-packed crystal the available free volume, $v_\mathrm{free}$, is that of the Voronoi cell (see Fig.~\ref{fig:curvedFV}). However, the true free volume is larger than this approximation, hence $C > 0$ for all lattices. Because its boundary is concave, i.e., $\frac{\partial^2 v_\mathrm{free}}{\partial x^2} < 0$, we also have that $\kappa_0 > 0$ for all lattices. No similar constraint, however, obviously fixes the sign of $\kappa_1$. (See Appendix \ref{sec:freeVolumeExpansion} for a fuller presentation.)

\begin{figure}[ht]
\includegraphics[width=0.9\linewidth]{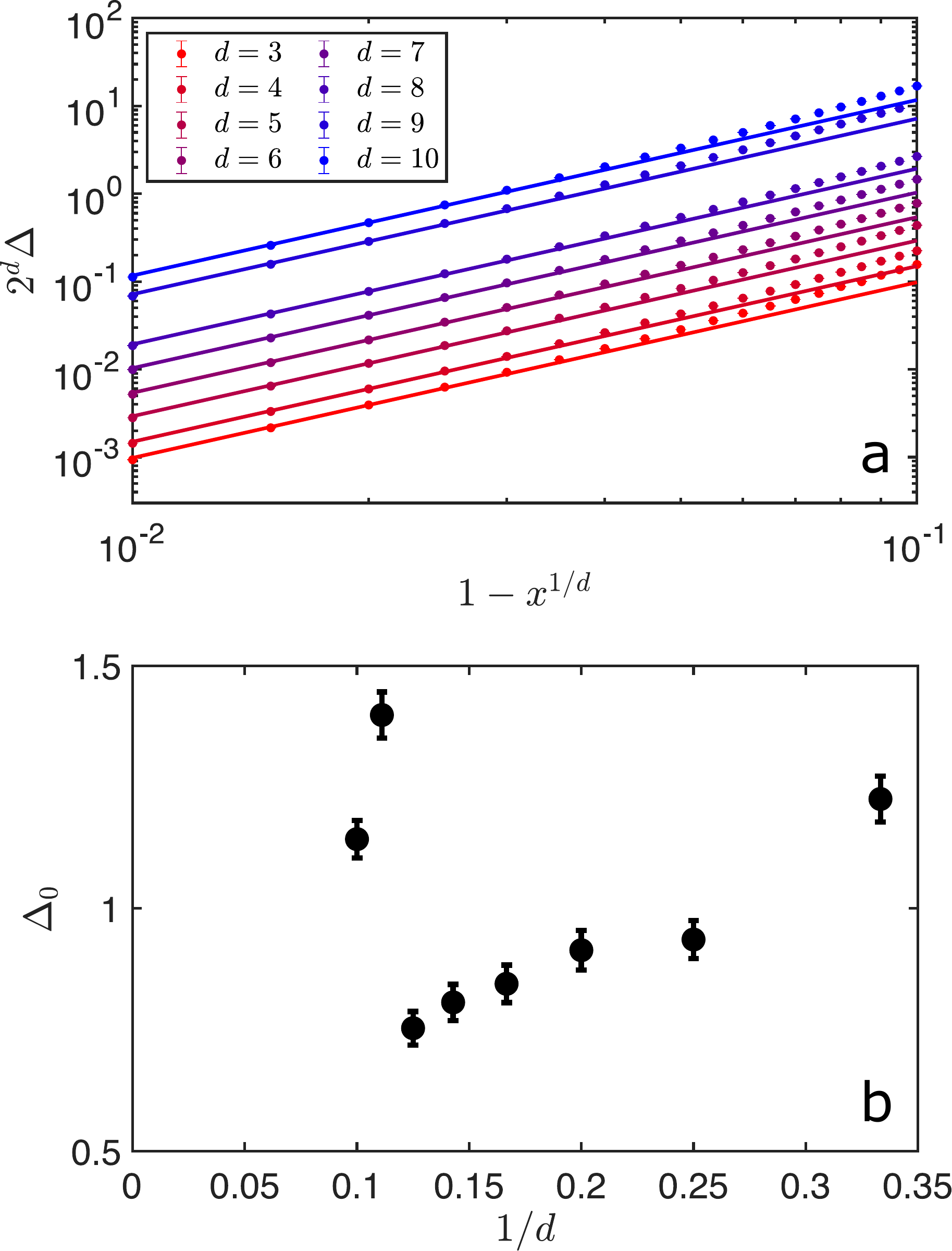}
\caption{\textbf{a)} Long-time MSD plateau for $d=3$-10 offset by a multiplicative factor of $2^d$ for visual clarity. \textbf{b)} From the scaling form in Eq.~\ref{eq:cagescaling} (solid lines), the prefactor $\Delta_0$ is extracted.}
\label{fig:deltaScaling}
\end{figure}

\subsection{Dynamical Cage size}
The cage size determined from the long-time limit of the mean squared displacement (MSD), $\langle r^2(t) \rangle$, can be used to estimate $a$ under simple assumptions. In order to compute the MSD, perfect crystals are first prepared by trivial planting. Equilibrium configurations are then sampled using the same Metropolis Monte Carlo scheme as in Ref.~\onlinecite{charbonneau_thermodynamic_2021}. 

A straightforward MSD computation is, however, inappropriate for $\lambda_9$, given that its nine internal soft modes permit unbounded motion along certain directions. For this crystal, the relevant MSD therefore excludes displacements along these soft dimensions,
\begin{equation}
\langle r^2(t) \rangle = \frac{1}{N} \sum_{i,\vartheta} [r^\vartheta_i(t) - r^\vartheta_i(0) - \Xi^\vartheta h^\vartheta_i - r^\vartheta_\mathrm{com}(t)]^2,
\end{equation}
where $\mathbf{r}_\mathrm{com}(t)$ denotes the center of mass of sublattice $\vartheta$ and $\{2\Xi^\vartheta\}$ denotes the (global) displacement vector along each mode. (The factor of $2$ is included for symmetrization.) Each of the nine sublattices contains half of all particles moving either in the positive or negative direction away from the center of mass of the sublattice. We denote the participation of each particle to each sublattice by the tensor elements $h_i^\vartheta=\pm 1$. The term $\Xi^\vartheta h_i^\vartheta$ then encodes the distance traveled by each particle from the center of mass of each sublattice during collective motions.

In all cases, the MSD is fitted to a stretched exponential
\begin{equation}
\langle r^2(t) \rangle = \Delta\big(1-e^{-(t/\tau_\beta)^{\gamma}}\big)
\label{eqn:tauDef}
\end{equation}
with relaxation time $\tau_\beta$, and stretching exponent, $\gamma < 1$, for time $t$ given in number of Monte Carlo sweeps, as it relaxes to its plateau height, $\Delta$. For the range of $d$, $N$, and $x$ considered, $\gamma$ typically varies between $0.4$ and $1$, and systematically increases with $d$ (Appendix~\ref{sec:equilibrationConstants}, Fig.~\ref{fig:equilibration}).  
The time constant depends only weakly on dimension and $x$--except for $d=3$ near coexistence---and $\tau_\beta$ are $\mathcal{O}(10)$. A system is deemed equilibrated after $t\geq 10\tau_\beta$, which can easily be achieved using standard computational resources.  Over $10,000$ independent snapshots for each $x$ and $d$ can thus be efficiently obtained.

As $x$ approaches unity, the typical linear cage size $\Delta$, scales as (Fig.~\ref{fig:deltaScaling})
\begin{equation}
\label{eq:cagescaling}
\Delta = \Delta_0(1-x^{1/d})^2,
\end{equation}
with prefactor $\Delta_0$. Using the partition function in Eq.~\eqref{eq:cryst} to compute the MSD of two random points in a $d$-dimensional sphere of radius $a$ separately gives
\begin{equation}
\Delta_0=2a^2\frac{d(d^2+2d-1)}{(1+d)^2(2+d)}.
\label{eq:dynamicA}
\end{equation}
Within this structural assumption, an estimate for $a$ can thus be extracted from the scaling of the MSD plateau height.

\begin{figure*}[htb]
\includegraphics[width=\linewidth]{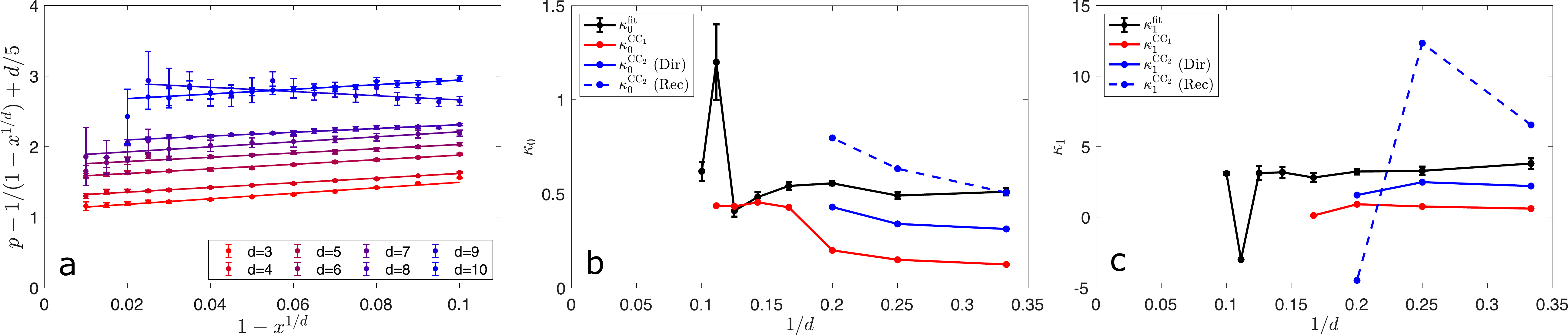}
\caption{\textbf{a)} Correction to the free volume equation of state in $d=3$-$10$. The constant and linear terms, $\kappa^\mathrm{fit}_0$ and $\kappa^\mathrm{fit}_1$, are estimated from a simple linear fit (lines) of the simulation results (points) over $1-x^{1/d} \in(0, 0.1)$.  The growth of the error bars, which denote 95\% confidence intervals, at high densities reflects the numerical difficulty of comparing two diverging quantities. Note that for visual clarity each curve is offset by $d/5$.  \textbf{b)} Comparison of $\kappa^\mathrm{fit}_0$ (black) from (a) with the value from the first order cell cluster expansion $\kappa^\mathrm{CC_1}_0$ (red) and second $\kappa^\mathrm{CC_1}_0$ (blue) using both expansions. As $d$ increases, the direct cluster expansion results appear to slowly converge towards the pressure calculation to first order, while the recursive expansion appears to diverge. \textbf{c)} Comparison of $\kappa^\mathrm{fit}_1$ from (a) with $\kappa^\mathrm{CC}_1$ and $\kappa_1^\mathrm{CC_2}$ using both cell cluster expansions. Here, the truncated recursive expansion oscillates as expected (see Appendix~\ref{sec:cellClusterAppendix}). In (b) and (c), lines are only provided as guides for the eye.}
\label{fig:speedyPDiff}
\end{figure*}

\subsection{Thermal integration}
An assumption-free estimate of $a$ can also be obtained by thermally integrating the crystal equation of state from a state of known entropy. Such high-accuracy entropies are available for $d=3$ up to close packing~\cite{speedy_pressure_1998}, but comparable results are limited to the liquid-crystal coexistence regime for $d=4$-$10$~\cite{van_meel_hard-sphere_2009,charbonneau_thermodynamic_2021}. For succinctness, we here briefly describe the integration scheme, and especially how it differs from that reported in Ref.~\onlinecite{charbonneau_thermodynamic_2021}.

The reduced pressure is first computed using the pair correlation at contact, $g(\sigma^+)$,
\begin{equation}
p = 1 + \frac{\overline{\varphi}}{2}g(\sigma^+) \ .
\end{equation}
These numerical results are then fitted using Eq.~\eqref{eqn:compressDeriv}
up to $\mathcal{O}(1-x^{1/d})$, which provides numerical estimates of $\kappa_0$ and $\kappa_1$. The fitted crystal equation of state captures simulation results well for all $d$ in this regime (Fig.~\ref{fig:speedyPDiff}), thus validating the form proposed by Rudd et al.~for expanding the free energy around close packing. Higher-order corrections would, however, be needed to describe pressures down to the fluid-crystal coexistence regime~\cite{charbonneau_thermodynamic_2021}. 

Absolute entropies at a reference density $x_0$ are computed by performing a Frenkel-Ladd integration at that state~\cite{frenkel_new_1984,polson_finite-size_2000} in all but $d=9$ where the periodic potential defined in Ref.~\onlinecite{charbonneau_thermodynamic_2021} is used instead. In order to optimize numerical accuracy, the reference state is taken near close packing, i.e., $x_0\approx 1$,  instead of near melting as in Ref.~\onlinecite{charbonneau_thermodynamic_2021}. However, a larger integration cutoff is then needed to prevent spheres from overlapping in the Einstein crystal limit of the Frenkel-Ladd scheme. These constraints are balanced by taking $x_0$ within the regime of validity of the fitted equation of state, but no denser. Because the reference entropy exhibits a significant size dependence--unlike the crystal equation of state--the thermodynamic $s_c(x_0)$ is further estimated using a standard finite-size scaling analysis (Appendix~\ref{sec:finSizeRef}).

A numerical estimate of $a$ can then be obtained via Eqs.~\eqref{eqn:rudd} and \eqref{eq:relationK0}-\eqref{eq:relationCA}, which in the limit $x\to 1$ yield
\begin{equation}
\begin{split}
d \ln a = s_c(x_0) - d\ln(1 - x_0^{1/d}) + \ln x_0 - 1 - \ln V_d \\ - d\kappa_0 (1-x_0^{1/d})
+ \frac{d}{2}(\kappa_0 + \kappa_1)(1-x_0^{1/d})^2 \ .
\label{eq:thermalIntA}
\end{split}
\end{equation}

\subsection{Summary of results from low-\texorpdfstring{$d$}{Lg} crystals}

\begin{figure}[ht]
%\vspace{0.5cm}
\includegraphics[width=0.9\columnwidth]{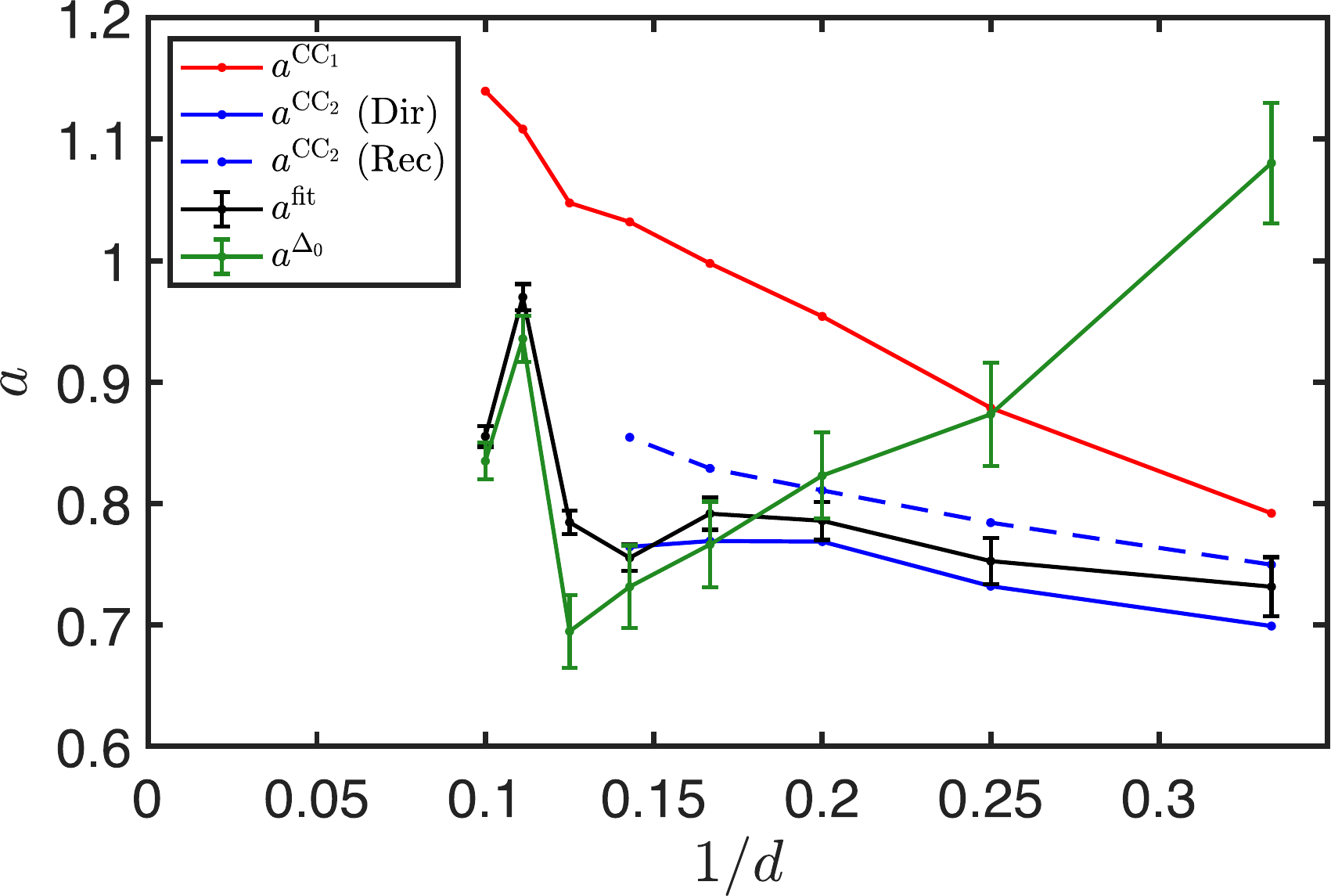}
\caption{Dimensional evolution of the lattice constant $a$ evaluated using three approaches. Cell-cluster expansion results are reported using Eq.~\eqref{eq:relationCA} to first ($a^\mathrm{CC_1}$ in red) and second ($a^\mathrm{CC_2}$ in blue) order for both direct (dashed line) and recursive (solid line) expansions (see Appendix~\ref{sec:cellClusterAppendix}). Thermal integration results, $a^\mathrm{fit}$ (black), from Eq.~\eqref{eq:thermalIntA} and dynamical estimates $a^{\Delta_0}$ (green) from Eq.~\eqref{eq:dynamicA} evolve qualitatively similarly. All estimates of $a$ suggest that $a \sim \mathcal{O}(1)$ as $d\rightarrow\infty$. Error bars denote a 95\% confidence interval; the cell cluster calculations are accurate to double precision truncation.}
\label{fig:tivscc}
\end{figure}

The cell-cluster expansion and the numerical estimates of $a$ from both the cage size and thermal integration are compared in Fig.~\ref{fig:tivscc}. The two numerical estimates, $a^{\Delta_0}$ and $a^\mathrm{fit}$, neatly converge as dimension increases. The spherical caging assumption of Eq.~\eqref{eq:dynamicA} on which the former relies, although fairly crude in low $d$, becomes increasingly inconsequential as $d$ increases. Going from the first level of the cell cluster expansion, $a^\mathrm{CC_1}$, to the second, $a^\mathrm{CC_2}$, also suggests a rapid convergence towards $a^\mathrm{fit}$ using both cell cluster expansions. Over the accessible $d$ range, however, the agreement does not markedly increase with dimension. Most importantly, all of these estimates support the conjecture $a \sim \mathcal{O}(1)$.

Because standard simulations of higher-dimensional systems become increasingly computationally challenging, were the expansion of Rudd et al.~fully controlled, it could help extend the present analysis. The convergence of the analytical expansion of $\kappa_0$ and $\kappa_1$ offers some hope in this direction, albeit only through the direct expansion strategy. Under the recursive strategy, the $\kappa_0$ and $\kappa_1$ terms  appear to diverge from the numerical results at second order, in support of Rudd et al.'s expectation that derivatives of this expansion might never converge in a truncated series~\cite{rudd_rigid_1968}. In this context, further formal expansion of the direct integration strategy to third order might be of interest. At the moment, however, the ability to numerically evaluate integrals of order $n$ in reasonable time is capped at $nd \approx 14$~\cite{koch_most_2005}.

Irrespective of any convergence concerns, the results for $\lambda_9$ stand out. In particular, thermal integration results for $\kappa_0$ and $a$ in $d=9$ are much larger than those of nearby dimensions, and $\kappa_1$ is of the opposite sign. These features largely track what one might expect of a crystal with soft modes. First, its (effective) cage should be elongated along soft directions, thus making $a$ larger. Second, because the free volume is elongated, its rate of increase with decreasing $x$, $-\frac{\partial^2 v_\mathrm{free}}{\partial x^2}$, should be larger than for standard caging, thus increasing $\kappa_0$. Third, the negative value of $\kappa_1$ might result from spheres in interlocking lattices being relatively less constrained and thus less likely to be in contact at high entropy points (where multiple soft modes are available). That said, unlike the crystal equation of state of other crystals, that of crystals with soft modes are expected to exhibit significant finite-size corrections, as discussed in Sec.~\ref{sec:conj}. Unfortunately, only a single system size is numerically available for this crystal~\cite{charbonneau_thermodynamic_2021}, and thus a systematic examination of these effects is not here feasible. Because the only path towards radical asphericity of the Voronoi cell--and thus significant deviations from $a \sim \mathcal{O}(1)$--is through the presence of a direction in which individual particles or subextensive collections of particles are not constrained or are only weakly constrained, a comparable analysis of lower-dimensional crystals containing soft modes, such as parallel hard cubes~\cite{swol_percolation_1987,jagla_melting_1998}, might thus be a more promising route to gain insight on this matter.

\section{Discussion and Conclusion}
\label{sec:final}

Under the assumptions of Sec.~\ref{sec:conj}, we summarize the three possible scenarios  available for high dimensional crystallization: (scenario A) if there is no crystallization, then the optimal packings are glasses; (scenario B) if crystallization occurs but only deep in the dynamically arrested region, then the densest crystal is only slightly more dense than the closest packed glass; (scenario C) if instead crystallization happens on the same scale as the dynamical arrest then the crystal close packing is $\maxphibar\sim e^{\a d}$, an exponential improvement over the Minkowski bound. Of these, scenario C relies on the constant $a$ being of order unity, while the others make no such requirement.

Through the use of both cell cluster expansions and numerical simulations of crystals in $d=3$-$10$, we have obtained three independent measures of $a$ that are roughly consistent with each other. % and consistent with $a \sim \mathcal{O}(1)$. 
We further observe that the crystal entropy is dominated by the free volume description in all $d$. Although we observe a significant polynomial correction that does not markedly decrease with increasing dimension, it does not significantly increase in crystals with soft modes either. It is therefore expected that these contributions remain subdominant to the free volume description in all $d$.

Three additional low-$d$ observations point to the relative likelihood of each of the three scenarios proffered. First, crystallization is thermodynamically favored at least up to $d=10$ and $\widehat{\varphi}_f \sim 1$, which is well below $\wh\varphi_d$. Second, $\alpha = \frac{1}{d}\ln{\overline{\varphi}_\mathrm{cp}}$ remains finite and approaches the zone of crystallization allowed by scenario C. Third, $a$ appears to remain $\mathcal{O}(1)$. Together, these results indicate that while all scenarios are possible, scenario C is most likely to be true, scenario B less likely, and scenario A less likely still. While future numerical simulations and analysis of higher dimensional crystals may be possible in a few additional dimensions, it seems improbable that such simulations would upend this ordering.

Finally, it should be noted that even if scenario C is correct, it might still be heavily kinetically suppressed. If crystallization proceeds via classical nucleation theory, in particular, then the competition between surface and volume terms in the free energy creates a barrier such that the nucleation time should scale exponentially with $d$; see, e.g., Ref.~\onlinecite[Eq.~(3.28)]{debenedetti_metastable_1996}
and Ref.~\onlinecite[Eq.~(8.15)]{parisi_theory_2020}. Although alternative crystallization schemes have been suggested in deeply supercooled liquids in $d=3$~\cite{filion_crystal_2010, sanz_crystallization_2011}, the geometrical peculiarities of three-dimensional space that underlie such mechanisms appear unlikely to find echo in any higher $d$.

\begin{acknowledgments}
We thank Yi Hu, Robert Hoy, and Henry Cohn for stimulating discussions. This work was supported by grants from the Simons Foundation (\#454937, Patrick Charbonneau, \#454955 Francesco Zamponi), the National Science Foundation (DMS-1847451, Will Perkins) and the European Research Council (ERC) under the European Union's Horizon 2020 research and innovation programme (grant agreement n. 723955 - GlassUniversality). The computations were carried out on the Duke Compute Cluster (DCC), for which the authors thank Tom Milledge’s assistance. Data relevant to this work have been archived and can be accessed at the Duke Digital Repository~\cite{data}.
\end{acknowledgments}

\appendix

\section{Cell-cluster expansion details}
\label{sec:cellClusterAppendix}

The expansion of crystal entropy in Eq.~\eqref{eqn:rudd} formally extends free volume theory. It was originally accompanied by a recursive cell-cluster expansion expressed as a series expansions for each of the constant terms ($C$, $D$, $E$, etc.). Each order $n$ of the expansion frees $n$ connected particles while keeping all others fixed. This allows a systematic calculation of the excess entropy added at each order. A derivation of these series can be found in Ref.~\onlinecite{rudd_rigid_1968}, and several terms in the $C$ and $D$ expansions have been computed for face-centered cubic and hexagonal close packed crystals in $d=3$~\cite{rudd_rigid_1968, koch_most_2005} and for triangular crystals in $d=2$~\cite{rudd_rigid_1968, stillinger_rigid_1965}. We here truncate the expansion to two-particle order, $n=2$. Because there is only one way to form a two-particle cluster in a Bravais-lattice, one does not need to separate and weigh contributions from different cell types, thus greatly simplifying the original formalism. (We also use a parameter expansion that corresponds to $C'$ and $D'$ in Ref.~\onlinecite{rudd_rigid_1968}.) We present in this Appendix only that which is necessary to perform these expansions.

The structure of the rest of the Appendix is as follows. In Sec.~\ref{sec:clusterDefinitions}, the recursive cell expansion of Rudd et al.~as well as our direct expansion are defined, and their relative benefits and drawbacks are discussed. In Sec.~\ref{sec:integralDefs} the integrals used to create both expansions are defined, and in Sec.~\ref{sec:intMethods} the strategy for computing these integrals is detailed. Finally, because the formalism hides many of the technical details, in Sec.~\ref{sec:IntegralCalculation}, we work out minimal examples for $d=2$.

\subsection{Cluster expansion series}
\label{sec:clusterDefinitions}

As noted in Sec.~\ref{sec:cellClusterMain}, the cell cluster expansion is not unique. We here detail two related such expansions: the recursive expansion taken from Ref.~\onlinecite{rudd_rigid_1968}, and the direct expansion. Both are equivalent to first cluster order, but differ significantly at higher orders.

\subsubsection{Recursive expansion}

In the recursive expansion, the terms $C$, $D$, and $E$ can be expanded through a series of cell cluster expansions, shown here to second order,
\begin{equation}
\begin{split}
C &=  \mathcal{C}_1 + \frac{\mathcal{N}_c}{2}\mathcal{C}_2 \\
D &= \mathcal{D}_1 + \frac{\mathcal{N}_c}{2}\mathcal{D}_2 \\
E &= \mathcal{E}_1 + \frac{\mathcal{N}_c}{2}\mathcal{E}_2
\label{eq:cdeExpansion}
\end{split}
\end{equation}
where $\mathcal{N}_c$ is the number of contacts per particle at close packing~\cite{conway_sphere_1993}. The order $n$ of expansions ($\mathcal{C}_n$, $\mathcal{D}_n$, and $\mathcal{E}_n$) indicates the number or particles allowed to move in the calculation, with all others being pinned. Higher order $n$ are thus expected---albeit not proven---to converge towards thermodynamic values.

The $\mathcal{C}$, $\mathcal{D}$, and $\mathcal{E}$ terms are computed by a recursive series of integrals $I^{(i)}_n$, which are defined in Sec.~\ref{sec:integralDefs},
\begin{equation}
\begin{split}
\mathcal{C}_1 &= -\ln I_1^{(0)}, \ \ \ \ \ \ \mathcal{C}_2 = -\ln I_2^{(0)} - 2 \mathcal{C}_1, 
\\
\mathcal{D}_1 &= -I_{1}^{(1)}/I_{1}^{(0)}, \ \ \ 
\mathcal{D}_2 = -I_{2}^{(1)}/I_{2}^{(0)} - 2\mathcal{D}_1,\\
\mathcal{E}_1 &=-I_{1}^{(2)}/I_{1}^{(0)} + \frac{1}{2}(I_{1}^{(1)}/I_{1}^{(0)})^2,\\
\mathcal{E}_2 &= -I_{2}^{(2)}/I_{2}^{(0)} + \frac{1}{2}(I_{2}^{(1)}/I_{2}^{(0)})^2 - 2\mathcal{E}_1.
\label{eq:intermediaryExpansions}
\end{split}
\end{equation}

The diverging behavior of the expansions for $D$ and $E$ at second order can readily be predicted in the form of Eq.~\eqref{eq:cdeExpansion}, which can be rewritten at that order as
\begin{equation}
\begin{split}
C &= (\mathcal{N}_c - 1)\ln{I_1^{(0)}} - \frac{\mathcal{N}_c}{2}\ln{I_2^{(0)}} \\
D &= (\mathcal{N}_c-1)\frac{I_1^{(1)}}{I_1^{(0)}} - \frac{\mathcal{N}_c}{2}\frac{I_2^{(1)}}{I_2^{(0)}} \\
E &= \bigg[(\mathcal{N}_c-1)\frac{I_1^{(2)}}{I_1^{(0)}} + \frac{\mathcal{N}_c}{4}\bigg(\frac{I_2^{(1)}}{I_2^{(0)}}\bigg)^2\bigg] \\ &-\bigg[ \frac{\mathcal{N}_c-1}{2}\bigg(\frac{I_1^{(1)}}{I_1^{(0)}}\bigg)^2 + \frac{\mathcal{N}_c}{2}\bigg(\frac{I_2^{(1)}}{I_2^{(0)}}\bigg)^2\bigg].
\end{split}
\end{equation}
In this form, it is evident that $D$ and $E$ are the difference of two large terms, and thus potentially oscillate in sign as a function of $d$. Higher-order terms will generically contain larger coefficients and are thus also expected to oscillate in sign for fixed $d$. By contrast, $C$ can be transformed into the log of a quotient of large terms, which generally converges more rapidly.

\subsubsection{Direct expansion}

Instead of creating a series in $C$, $D$ and $E$, which explicitly presents a dependence on lower orders, we can directly expand the integrals $I_n^{(i)}$ defined in Sec.~\ref{sec:integralDefs}, noting that for Bravais-lattice packings there is only one type of connected cluster at both first and second order. Thus, any direct expansion finds that individual particle motion is a subset of the motion of two connected particles, and is thus contained entirely in the second-order integral. The same cannot be said of third-order expansions, which have multiple forms in every lattice (aside from the trivial case of $d=1$) that must be weighted appropriately by their contributions. Such third-order terms are not further considered here.

The form of the direct expansion is identical to the recursive expansion to first order. However, to second order, the equations for the expansion coefficients should not attempt to isolate the motion unique to the pair by subtracting out contributions to the pair from individual particle motion. A caveat is that while $D$ and $E$ integrals explicitly avoid double counting at all orders, the $C$ integral under the direct expansion must explicitly add it in the form of a factor of $2$ for each dimension. The expressions then become
\begin{equation}
\begin{split}
C &= -\ln \big(2^{-d}I_2^{(2)}\big) \\
D &= -I_2^{(1)}/I_2^{(0)} \\
E &= -I_{2}^{(2)}/I_{2}^{(0)} + \frac{1}{2}(I_{2}^{(1)}/I_{2}^{(0)})^2.
\end{split}
\end{equation}
This set of equations doesn't contain a difference of large numbers, and is thus potentially better behaved. For this reason, the direct expansion for both $D$ and $E$ is used in the main text.

\subsection{Integral definitions}
\label{sec:integralDefs}
Here, $I_{n}^{(i)}$ integrals provide two expansions: the subscript $i$ refers to which of the terms ($C$, $D$, $E$, etc.) is modified, and $n$ gives the number of free particles. $I_{n}^{(i)}$ is thus an $nd$-dimensional integral. Contributions of the cell clusters are grouped, such that only connected clusters with $n$ particles contribute at $n$-th order. For lattice packings, only one integral is necessary for $n=1$ and $n=2$. For $n>2$, all crystals and packings have a variety of cluster configurations which must be catalogued~\cite{stillinger_rigid_1965, salsburg_rigid_1967, rudd_rigid_1968, koch_most_2005}, while non-Bravais-lattice packings may have multiple terms at order $n=2$ (see, e.g., the treatment of hexagonal close packing~\cite{rudd_rigid_1968}), or even at order $n=1$ (eg. binary code based packings for which the number of neighbors is nonuniform:  $P_{9a}$, $P_{10a}$, $P_{10b}$, etc.~\cite{conway_sphere_1993}). 

These integrals are
\begin{align}
\label{eqn:cIntegral} 
I_n^{(0)} &=\int_\mathcal{R} d\Omega \\
I_n^{(1)} &= \int_\mathcal{R} \bigg[\sum_{i<j}p_{ij}^{(1)}\delta(1+\mathbf{w}_{ij}\cdot\mathbf{z}_{ij})\bigg]d\Omega
\end{align} 
%\end{equation}
%
%\begin{equation}
%\begin{split}
\begin{widetext}
\begin{equation}
I_n^{(2)} = \!\!\!\int_\mathcal{R}\!\!\!\bigg( \sum_{i<j} \big[2p_{ij}^{(2)}\delta(1+\mathbf{w}_{ij}\cdot\mathbf{z}_{ij}) + (p_{ij}^{(1)})^2 \delta'(1+\mathbf{w}_{ij}\cdot\mathbf{z}_{ij}) \big] + 
\!\!\!\!\!\sum_{\substack{i<j\\k<\ell \neq i,j}}\!\!\!\!\!%\sum_{}
\big[p_{ij}^{(1)}p_{k\ell}^{(1)}\delta(1+\mathbf{w}_{ij}\cdot\mathbf{z}_{ij})\delta(1+\mathbf{w}_{k\ell}\cdot\mathbf{z}_{k\ell})\big]\bigg)d\Omega
%\end{split}
\label{eqn:eIntegral}
%\end{equation}
\end{equation}
\end{widetext}
where $\mathbf{z}_i = \frac{1}{\rho^{1/d}_{cp}\sigma(1-x^{1/d})} (\mathbf{y}_i - \mathbf{y}_i^{(0)})$ is the scaled displacement of particle $i$ from its lattice position $\mathbf{y}^{(0)}_i$, ${\mathbf{z}_{ij} = \mathbf{z}_j - \mathbf{z}_i}$ (and likewise with $\mathbf{y}_{ij}$ and $\mathbf{y}^{(0)}_{ij}$), ${\mathbf{w}_{ij} = \mathbf{y}_{ij}^{(0)}/(\maxrho^{1/d}\sigma)}$ is the unit vector between nearest neighbors $i$ and $j$, $\Theta(x)$ is the Heaviside function, $\delta(x)$ is the Dirac delta function, $\delta'(x)$ is the derivative of the Dirac delta function, $\mathcal{R}$ is the total volume of $nd$-dimensional space, $p_{ij}^{(1)} = \frac{1}{2}[z_{ij}^2 - (\mathbf{w}_{ij}\cdot\mathbf{z}_{ij})^2]$, $p_{ij}^{(2)} = p_{ij}^{(1)}(\mathbf{w}_{ij}\cdot\mathbf{z}_{ij})$, and $d\Omega = \prod_{k<\ell} \Theta(1 + \mathbf{w}_{k\ell} \cdot \mathbf{z}_{k\ell}) d\mathbf{z}_1 \dotsi d\mathbf{z}_n$. Note that $\mathbf{w}_{ij}$ and $\mathbf{z}_{ij}$ are $d$ dimensional vectors of $nd$ variables.

In order to evaluate $I_n^{(1)}$ and $I_n^{(2)}$, the following two delta function identities are particularly helpful,
\begin{equation}
\begin{split}
\int^\infty_{-\infty}f(x,\{y_i\})\delta(a+bx+\mathbf{c}\cdot\mathbf{y})dx \\ = \frac{1}{|b|}f\bigg(-\frac{a+\mathbf{c}\cdot\mathbf{y}}{b},\{y_i\}\bigg)
\label{eqn:deltaID}
\end{split}
\end{equation}
\begin{equation}
\begin{split}
\int^\infty_{-\infty}f(x,\{y_i\})\delta'(a+bx+\mathbf{c}\cdot\mathbf{y})dx \\= -\frac{1}{|b|}\frac{\partial}{\partial x} f(x,\{y_i\})\bigg|_{x=-\frac{a+\mathbf{c}\cdot\mathbf{y}}{b}}
\label{eqn:deltaPrimeID}
\end{split}
\end{equation}
where $\mathbf{y}$ is a spatial vector with elements $\{y_i\}$, $\mathbf{c}$ is a constant vector with elements $\{c_i\}$, and ${a,b,c_i \in \mathbb{R}}$ for all $i$. When performed over finite ranges, these integrals are zero unless the range includes the value ${x=-(a+\mathbf{c}\cdot\mathbf{y})/b}$. Once any delta functions is evaluated, all $I_n^{(i)}$ are reduced to (a sum of) integrals of polynomials over spaces bounded by a set of planes, which are thus lower dimensional polytopes. This construction immediately invokes the half-space representation (H-Representation)~\cite{ziegler_convex_2003} of $(nd-\delta_n)$-dimensional polytopes, where $\delta_n$ is the number of delta functions evaluated on a given integral. The simplest terms, $I^{(0)}_n$ (and thus $C$), quantify the volume of these polytopes, and thus the free volume available to a given particle subject to the motion of its neighbors. (A visual representation of $I^{(0)}_2$ in $d=2$ is provided in Fig.~\ref{fig:stillDiagram}a.) $I^{(1)}_n$ and $I^{(2)}_n$ (and thus $D$ and $E$) terms involve integrating a polynomial over the surface and edges of the bounding polytope, respectively, 
and thus describe the cell curvature and torsion ~\cite{rudd_rigid_1968}. Numerical values for the cell-cluster integrals are provided in Table~\ref{table:integralTable}.

It is important to note that any term integrated over a domain with dimension $< nd-\delta_n$ is necessarily zero. For example, the final term for $I^{(2)}_1$ in $d=3$ is a one-dimensional integral and contains terms whose domains are $0$-dimensional points. Likewise, the final term for $I^{(2)}_2$ in $d=3$ is a four-dimensional integral, but contains terms whose domains may be $0$, $1$, $2$, or $3$ dimensional. All of these lower-dimensional terms evaluate to zero.

\subsection{Computing cell-cluster integrals}
\label{sec:intMethods}

Because $I_1^{(0)}$ is a simple function of the Wigner-Seitz cell volume, which can straightforwardly be looked up~\cite{conway_sphere_1993}, it can be calculated as 
\begin{equation}
I_1^{(0)} = \frac{\maxrho \sigma^d}{2^d}.
\label{eq:c1exact}
\end{equation}
Note that $\rho_\mathrm{cp}$ is here measured with respect to the particle diameter $\sigma$, whereas Ref.~\onlinecite{conway_sphere_1993} used a radius convention. The conversion simply entails rewriting Eq.~\eqref{eq:c1exact} as $I_1^{(0)} =  \tilde{\rho}_\mathrm{cp} (\sigma/2)^d$, whereupon $\tilde{\rho}_\mathrm{cp}$ is measured with respect to radii and $\sigma/2$ sets the unit of length.

\begin{figure}[b]
\includegraphics[width=0.9\linewidth]{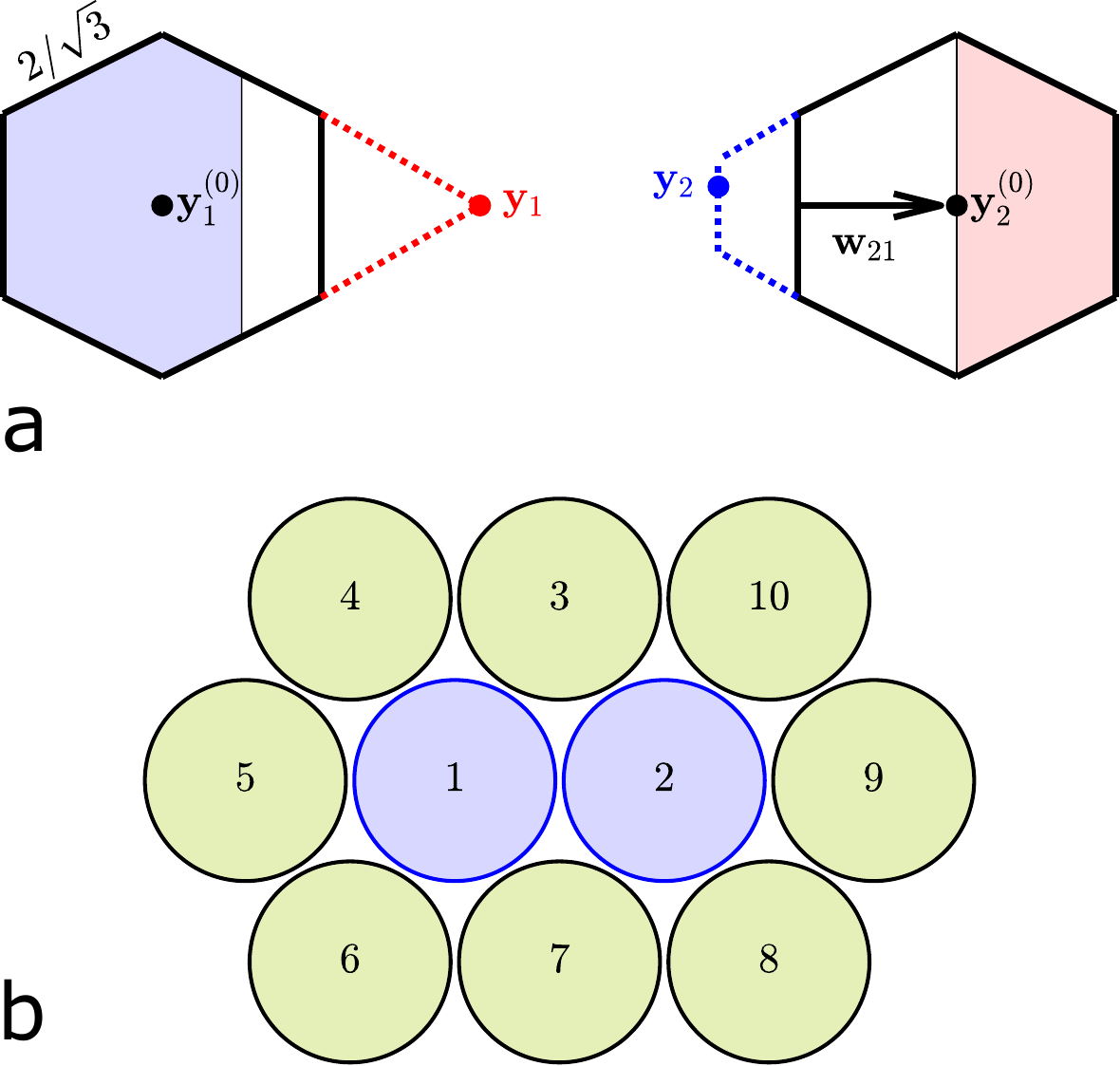}
\caption{\textbf{a)} Diagram of the boundaries used in the $I^{(i)}_2$ calculations in $d=2$, adapted from Ref.~\onlinecite{stillinger_rigid_1965}. Lattice positions for points $1$ and $2$ are given by $\mathbf{y}^{(0)}_1$ and $\mathbf{y}^{(0)}_2$, respectively, and $\mathbf{w}_{21}$ is shown emanating from its associated boundary. Black hexagons represent the free volume available to each particle if the other is fixed. (The construction is enlarged for clarity; in the calculation $x \to 1$, which makes the free volume infinitesimal). Two hypothetical perturbations $\mathbf{y}_1$ and $\mathbf{y}_2$ are shown to illustrate the excluded free volume. If $1$ is at $\mathbf{y}_1$, then $2$ is restricted to the shaded red area and likewise, if $2$ is in the shaded red region, $1$ is allowed anywhere in its hexagon or the extended dashed red region. The same can be said of the less extreme perturbation shown in blue. \textbf{b)} Diagram labeling the spheres in \textbf{a} that are allowed to move in $I^{(i)}_2$ calculations ($1$ and $2$), and those that are fixed ($3$-$10$).}
\label{fig:stillDiagram}
\end{figure}

All other $I_n^{(i)}$ are computed using the LattE package~\cite{baldoni_users_2013}, which takes as inputs the H-Representation of the boundary and the polynomial integrand. The case, $I_1^{(1)}$, is a sum of integrals each projected by a delta function onto the planes $1+\mathbf{w}_{ij}\cdot\mathbf{z}_{ij} = 0$. Here, because all particles but one are fixed, $\mathbf{z}_{ij} = \mathbf{z}_1$. The integrals then simplify to $(d-1)$-dimensional integrals with boundaries set by the $(d-2)$-dimensional surfaces defined by the intersection of planes.

The case $I_2^{(0)}$ unpins a second particle, which, for Bravais-lattices, has the same set of neighbor vectors $\mathbf{w}_{ij}$ and integrates over a simple unit polynomial, $1$. While the neighbor vectors are the same, the coordinated motion of the two particles encoded in the $2d$-dimensional vector $\mathbf{z}$ yields a set of inequalities (the case $d=2$ is shown explicitly in Sec.~\ref{sec:IntegralCalculation}) that can be given in the H-Representation as 
\begin{equation}
-A\cdot \mathbf{z} \le \mathbf{1}
\label{eq:AMatrix}
\end{equation}
where $A$ is a $(2\mathcal{N}-1) \times 2d$ matrix and $\mathbf{1}$ represents a $(2\mathcal{N}-1)$ dimensional column vector. If we choose $\mathbf{w}_{12}$ to align with the $x$-direction (as in Fig.~\ref{fig:stillDiagram}), then $A$ can be written strictly in terms of the set of vectors $\{\mathbf{w}_{1,j\neq 2}\}$ and $\{\mathbf{w}_{2,i\neq 1}\}$, for which the rows of $A$ are elements of the set $\{\{1,0,\shortdots{},-1,0,\shortdots{}\},\allowbreak \{\mathbf{w}_{1,j\neq 2},0,\shortdots{}\},\allowbreak \{0,\shortdots{},\mathbf{w}_{2,i\neq 2}\}\}$. In this example $j=3$-$7$ and $i=3,7$-$10$. The condition involving particles $1$ and $2$ is then given by the first term in the set. This construction leads to a straightforward---albeit nontrivial---calculation scheme for $I_n^{(0)}$, which simply integrates over the $2d$-dimensional volume defined by the set of Eq.~\eqref{eq:AMatrix}. In practice, however, the time required to evaluate the $2d$-dimensional integrals grows steeply with $d$. Leveraging rotation and reflection symmetries is thus computationally key. Such symmetries comprise the point group of $A$ and a subset of them are identified by applying $2d \times 2d$ rotation matrices $R$ with normal vector $\tilde{\mathbf{R}}$ about a hyperplane ${1+\tilde{\mathbf{R}}\cdot\mathbf{z}_{ij}=0}$ such that
\begin{equation}
A^T = (R)^m \cdot A^T,
\end{equation}
where the superscript $T$ represents the transpose, and the set of columns remains unchanged under $\{R\cdot A^T\} = \{A^T\}$, signifying an $m$-fold symmetry. The $R$ studied here involve only simple rotations of the form 
\begin{equation}
\begin{split}
R_{ij}(k,\ell\neq k,m) &= \delta_{ij}\big(1 + \delta_{ik}\delta_{j\ell}[\cos(2\pi/m)-1]\big) \\&+ (\delta_{i\ell}\delta_{jk}-\delta_{ik}\delta_{j\ell})\sin(2\pi/m)
\end{split}
\end{equation}
and the symmetries are included in the H-Representation as $z_k \ge 0$ and $\cos(2\pi/m)z_k + \sin(2\pi/m)z_\ell \ge 0$, with each $m$-fold symmetry providing a speedup factor of $m$. If used, each (non-redundant) $m$-fold symmetry multiplies the result of the integral by a factor $m$.

We also employ reflection symmetries, for which one needs the set $\{\mathfrak{R}^{(k)}\}$ that operate on the set of unit vectors $\{\hat{e}_i\}$ as ${\mathfrak{R}^{(k)}\hat{e}_i = (1-2\delta_{ik})\hat{e}_i}$ flipping the sign of only elements in the $k$ position, but yielding ${A^T = (\mathfrak{R}^{(k)})^2 \cdot A^T}$ and ${\{\mathfrak{R}^{(k)}\cdot A^T\} = \{A^T\}}$. The matrix elements of the transformation are then given by
\begin{equation}
\mathfrak{R}^{(k)}_{ij} = \delta_{ij}(1-2\delta_{ik}),
\end{equation}
and the constraint added to the H-Representation is $z_{k} \ge 0$. If used, each (non-redundant) reflection symmetry speeds up the calculation by a factor of $2$ and accounts for a multiplicative factor of $2$ to the reduced integral.

(This approach is successful for $d\le6$, but results for $d=7$ took three weeks to obtain and may be unreliable. A more generic scheme to determine these rotations might improve the situation. In general, one need not find the entire point group of $A$; its subset of rotations larger than the set of simple rotations suffices. If the point group of $\mathbf{w}_{1,j}$ is known, then a subset of $\mathbf{w}_{i,j}$ can also be generated.)

The integral $I_2^{(1)}$ builds off of the formulation of $I_1^{(1)}$, but uses the geometry shown in Fig.~\ref{fig:stillDiagram}a. Because elements of $\mathbf{z}_{i}$ are only nonzero when particle $i$ is unpinned, the polynomial $p^{(1)}_{ij}$ only contains terms with elements of either $\mathbf{z}_1$ or $\mathbf{z}_2$, except in the case of the surface defined by $\mathbf{w}_{12}$. On the surface defined by $\mathbf{w}_{12}$, we have ${z_{ij}^2 = (\mathbf{z_2} - \mathbf{z}_1)^2}$ and $\mathbf{w}_{ij}\cdot\mathbf{z}_{ij} = z_j^x - z_i^x$, where $z_j^x$ denotes the $x$-component of the displacement of particle $j$ from its reference position. Each of these integrals contains a delta function, which is manually evaluated via Eq.~\eqref{eqn:deltaID}. The result is an integral of a polynomial over a $(2d-1)$-dimensional polytope. Here again symmetries can be leveraged, though to a lesser degree, because they must keep the polytope and the polynomial invariant. We thus restrict our consideration to reflection symmetries $\{\mathcal{R}^{(k)}\}$, for which the polynomial $f$ is even in the $k$-component, ie. $f(-x^{(k)},\{x^{(i\neq k)}\}) = f(x^{(k)},\{x^{(i\neq k)}\})$.

Although the integrals $I_1^{(2)}$ and $I_2^{(2)}$ involve significantly more terms, they follow directly from the calculations of $I_1^{(1)}$ and $I_2^{(1)}$, after applying the identities in Eqs.~\eqref{eqn:deltaID} and \eqref{eqn:deltaPrimeID}. There are, however, two caveats. First, Eq.~\eqref{eqn:deltaPrimeID} involves taking a partial derivative of the function $(p_{ij}^{(1)})^2$ with respect to the variable being manually integrated via the delta function. The result is nevertheless a polynomial that can be integrated as usual. Second, two delta functions must be evaluated in the second sum. Hence, for $I_1^{(2)}$ the first sum of Eq.~\eqref{eqn:eIntegral} is an integral over a $(d-1)$-dimensional polytope and the second sum is an integral over a $(d-2)$-dimensional polytope. Meanwhile for $I_2^{(2)}$ the first sum of Eq.~\eqref{eqn:eIntegral} is an integral over a $(2d-1)$-dimensional polytope and the second sum is an integral over a $(2d-2)$-dimensional polytope.

\begin{table*}[htb]\centering
\begin{tabular}{c | c | c | c | c | c | c }
\hline
\hline
crystal & $I^{(0)}_1$ & $I^{(0)}_2$ & $I^{(1)}_1$ & $I^{(1)}_2$ & $I^{(2)}_1$ & $I^{(2)}_2$ \\
\hline
$D_3$ & $4\sqrt{2}$ & 467/15 & $3\sqrt{2}/2$ & 1318/45 & $47\sqrt{2}/10$ & 29493/224 \\
$D_4$ & 8 & 1294/21 & 24/5 & 164372/945 & 16 & 382931/945\\
$D_5$ & $8\sqrt{2}$ & 38713/315 & $8\sqrt{2}$ & 508482/1925 & $1479\sqrt{2}/56$ & 898.8448849657\\
$E_6$ & $8\sqrt{3}$ & 186.1604795252 & $144\sqrt{3}/7$ & -- & 68.7308002580 & --\\
$E_7$ & 16 & -- & 51 & -- & -- & --\\
$E_8$ & 16 & -- & 4496/81 & -- & -- & --\\
$\lambda_9$ & $16\sqrt{2} $ & -- & 88.9850541943 & -- & -- & --\\
$P_{10c}$ & 128/5 & -- & -- & -- & -- & --\\
\hline
\hline
\end{tabular}
\caption{\textbf{Cell cluster integrals for each crystal.} Symbolic forms are given when known. Otherwise, values are exact to machine precision but are rounded to ten decimal places.}
\label{table:integralTable}
\end{table*}

\subsection{Calculation of cluster integrals in \texorpdfstring{$d=2$}{Lg}}
\label{sec:IntegralCalculation}

To the best of our knowledge, the only worked out examples in the literature for any of the cell cluster integrals are $I^{(0)}_2$ and two of the three configurations of $I^{(0)}_3$ in $d=2$  ~\cite{stillinger_rigid_1965}. Unfortunately, the methods used to evaluate these integrals are not easily generalizable to higher dimension as they rely on a set of special identities. There are also several errors in the original calculations of Rudd et al.~\cite{rudd_rigid_1968}, such as (but not limited to): (i) the calculations of $I^{(1)}_2$ in both $d=2$ and $d=3$; (ii) the linear configuration of $I^{(1)}_3$ in $d=2$ (although the other two configurations associated with $I^{(1)}_3$ appear to be correct); and (iii) several of the $I^{(0)}_n$ contributions for the hexagonal close packing in $d=3$ for $n \ge 2$ (as previously noted in Ref.~\onlinecite{koch_most_2005}). Because of these issues, and because the extension to $I^{(1)}_n$ from $I^{(0)}_n$ is non-trivial, we feel it is helpful to provide a more extended set of examples. We thus here explicitly calculate $I^{(0)}_1$, $I^{(1)}_1$, $I^{(0)}_2$, and $I^{(1)}_2$ in $d=2$, whose values are reported---some incorrectly, as noted---in Refs.~\cite{rudd_rigid_1968,stillinger_rigid_1965,salsburg_rigid_1967}. From these results, it is trivial to extend the calculation scheme to $I^{(2)}_n$ using the methods of Sec.~\ref{sec:intMethods}.

From Eq.~\eqref{eq:c1exact} and $\maxrho \sigma^d = 2/\sqrt{3}$, we find that $I^{(0)}_1 = 2\sqrt{3}$. Using the orientation of Fig.~\ref{fig:stillDiagram}b (with particle $2$ fixed), the neighbor vectors are then $\{\mathbf{w}_{1,i}\} = \allowbreak \{1,0\}, \allowbreak \{-1,0\}, \allowbreak  \{\frac{1}{2}, \frac{\sqrt{3}}{2}\}, \allowbreak  \{\frac{1}{2}, -\frac{\sqrt{3}}{2}\}, \{-\frac{1}{2},\frac{\sqrt{3}}{2}\}, \allowbreak  \{-\frac{1}{2},-\frac{\sqrt{3}}{2}\}$. This set forms a regular polytope, and so $I^{(1)}_1$ consists of a sum of $6$ identical integrals. Furthermore, because all other particles are pinned, $z_j = 0$ for all $j\neq 1$, and thus  $z_{ij} = z_i$. The integral associated with the face at $\mathbf{w}_{1,2}=\{1,0\}$ then gives
\begin{equation}
\begin{split}
I^{(1)}_1 &= 6\int_\mathcal{R} \frac{1}{2} [z^2_{1,2} - (\mathbf{w}_{1,2}\cdot\mathbf{z}_{1,2})^2 ]\delta(1+\mathbf{w}_{1,2}\cdot\mathbf{z}_{1,2})d\Omega\\
&= 3\int_{-1/\sqrt{3}}^{1/\sqrt{3}} \big[(x^2 + y^2) - x^2 \big]dy = \frac{2}{3\sqrt{3}}.
\end{split}
\end{equation}
For $I^{(0)}_2$, we label the (fixed) particles $j=3$-$10$ as in Fig.~\ref{fig:stillDiagram}b, using Fig.~\ref{fig:stillDiagram} to write the integral (with $x$- and $y$-coordinates of $\mathbf{z}_i$ as $z_i^x$ and $z_i^y$ and the $x$- and $y$-components of $\mathbf{z}_{ij}$ as $z_{i,j}^x$ and $z_{i,j}^y$):
\begin{equation}
\begin{split}
I^{(0)}_2 &= \int_\mathcal{R}d\mathbf{z}_1 d\mathbf{z}_2 H(1-\mathbf{w}_{1,3}\cdot\mathbf{z}_{1,3}) H(1-\mathbf{w}_{1,4}\cdot\mathbf{z}_{1,4})\\
&\times H(1-\mathbf{w}_{1,5}\cdot\mathbf{z}_{1,5}) H(1-\mathbf{w}_{1,6}\cdot\mathbf{z}_{1,6})H(1-\mathbf{w}_{1,7}\cdot\mathbf{z}_{1,7})\\ &\times H(1-\mathbf{w}_{2,7}\cdot\mathbf{z}_{2,7}) H(1-\mathbf{w}_{2,8}\cdot\mathbf{z}_{2,8}) H(1-\mathbf{w}_{2,9}\cdot\mathbf{z}_{2,9})\\&\times H(1-\mathbf{w}_{2,10}\cdot\mathbf{z}_{2,10}) H(1-\mathbf{w}_{2,3}\cdot\mathbf{z}_{2,3})\\ &=\int_\mathcal{R}dz_1^x dz_1^y dz_2^x dz_2^y H(1-z_1^x+z_2^x) H(1+z_1^x) H(1-z_2^x)\\ &\times H(1-\frac{1}{2}z_1^x -\frac{\sqrt{3}}{2}z_1^y) H(1-\frac{1}{2}z_1^x +\frac{\sqrt{3}}{2}z_1^y)\\
&\times H(1+\frac{1}{2}z_1^x -\frac{\sqrt{3}}{2}z_1^y) H(1+\frac{1}{2}z_1^x +\frac{\sqrt{3}}{2}z_1^y) \\
&\times H(1-\frac{1}{2}z_2^x -\frac{\sqrt{3}}{2}2_1^y) H(1-\frac{1}{2}z_2^x +\frac{\sqrt{3}}{2}2_1^y) \\ &\times H(1+\frac{1}{2}z_2^x -\frac{\sqrt{3}}{2}2_1^y)H(1+\frac{1}{2}z_2^x +\frac{\sqrt{3}}{2}2_1^y).
\end{split}
\label{eq:c2ExampleInt}
\end{equation}
With this construction, Eq.~\eqref{eq:AMatrix} can be rewritten as the following system of inequalities:
\begin{equation}
\begin{bmatrix}
1 & 0 & -1 & 0 \\
-1 & 0 & 0 & 0 \\
0 & 0 & 1 & 0 \\
1/2 & \sqrt{3}/2 & 0 & 0 \\
1/2 & -\sqrt{3}/2 & 0 & 0 \\
-1/2 & \sqrt{3}/2 & 0 & 0 \\
-1/2 & -\sqrt{3}2 & 0 & 0 \\
0 & 0 & 1/2 & \sqrt{3}/2 \\
0 & 0 & 1/2 & -\sqrt{3}/2 \\
0 & 0 & -1/2 & \sqrt{3}/2 \\
0 & 0 & -1/2 & -\sqrt{3}/2 \\
\end{bmatrix} \cdot
\begin{bmatrix}
z_1^x \\ z_1^y \\ z_2^x \\ z_2^y
\end{bmatrix}
\leq
\begin{bmatrix}
1\\ 1\\ 1\\ 1\\ 1\\ 1\\ 1\\ 1\\ 1\\ 1\\ 1\\
\end{bmatrix}.
\label{eq:c2exampleMatrix}
\end{equation}
The resulting integral can be evaluated in a variety of ways. (See, e.g., Ref.~\onlinecite[Eqs.~33-37]{stillinger_rigid_1965} and Refs.~\onlinecite{rudd_efficient_1985,rudd_methods_1970} for a general treatment.) As these approaches quickly become unwieldy, we here use the LattE package, which has been developed precisely for computing this type of integrals and operates on the H-Representation~\cite{baldoni_users_2013}. Note that two reflection symmetries (or rotations by $\pi/2$) exist, $z_1^y \ge 0$ and $z_2^y \ge 0$, and thus the calculation can be sped up by a factor of four by adding two rows to Eq.~\eqref{eq:c2exampleMatrix},
\begin{equation}
\begin{bmatrix}
0 & -1 & 0 & 0 \\
0 & 0 & 0 & -1 \\
\end{bmatrix}\cdot
\begin{bmatrix}
z_1^x \\ z_1^y \\ z_2^x \\ z_2^y
\end{bmatrix}
\leq
\begin{bmatrix}
0\\ 0
\end{bmatrix}.
\end{equation}
and the reduced integral must be multiplied by a factor of four as well. Note also that if these reflection constraints are applied, rows 5, 7, 9, and 11 of Eq.~\eqref{eq:c2exampleMatrix} are redundant. The calculation, whether using symmetries or not, gives $I^{(0)}_2 = \ln(216/217)$.

To calculate $I^{(1)}_2$, we simply build off of the calculation of $I^{(0)}_2$, which gives an explicit form for $d\Omega$. In total, calculating $I^{(1)}_2$ requires summing $11$ integrals, which all follow the same scheme. We here provide a generic case, the integral of $p_{12}^{(1)}$, by calculating the polynomial $f$ and the bounding $3$-planes. We first note that 
\begin{equation}
\begin{split}
p_{12}^{(1)} &= \frac{1}{2}\big(\big[(z_2^x - z_1^x)^2 + (z_2^y - z_1^y)^2\big] - (z_2^x - z_1^x)^2\big) \\&= \frac{1}{2}(z_2^y - z_1^y)^2.
\end{split}
\end{equation} 
Using Eq.~\eqref{eqn:deltaID} to integrate over $z_1^x$ and equating ${z_1^x = 1+z_2^x}$ in the limits, the integral then becomes
\begin{equation}
\begin{split}
&\int_\mathcal{R} p_{12}^{(1)}\delta(1-z_1^x+z_2^x)d\Omega = \\  &\int_\mathcal{R} dz_1^y dz_2^x dz_2^y \bigg[\frac{1}{2} (z_2^y - z_1^y)^2\bigg] H(2-z_2^x) H(1+z_2^x) \\ &\times H(\frac{1}{2}+\frac{\sqrt{3}}{2}z_1^y+\frac{1}{2}z_2^x) H(\frac{3}{2}+\frac{\sqrt{3}}{2}z_1^y-\frac{1}{2}z_2^x)\\ &\times H(\frac{1}{2}-\frac{\sqrt{3}}{2}z_1^y+\frac{1}{2}z_2^x) H(\frac{3}{2}-\frac{\sqrt{3}}{2}z_1^y-\frac{1}{2}z_2^x)  \\ &\times H(1+\frac{1}{2}z_2^x+\frac{\sqrt{3}}{2}z_2^y) H(1+\frac{1}{2}z_2^x-\frac{\sqrt{3}}{2}z_2^y)\\ &\times
H(1-\frac{1}{2}z_2^x+\frac{\sqrt{3}}{2}z_2^y)
H(1-\frac{1}{2}z_2^x-\frac{\sqrt{3}}{2}z_2^y).
\end{split}
\end{equation}
Because the integrand itself doesn't have any $z_1^x$ dependence to be substituted, this polynomial form can be integrated over a polytope given by the following H-Representation:
\begin{equation}
\begin{bmatrix}
1 & 0 & 0 \\
0 & -1 & 0 \\
-\sqrt{3}/2 & -1/2 & 0 \\
-\sqrt{3}/2 & 1/2 & 0 \\
\sqrt{3}/2 & -1/2 & 0 \\
\sqrt{3}/2 & 1/2 & 0 \\
0 & -1/2 & -\sqrt{3}/2 \\
0 & -1/2 & \sqrt{3}/2 \\
0 & 1/2 & -\sqrt{3}/2 \\
0 & 1/2 & \sqrt{3}/2 \\
\end{bmatrix}\cdot
\begin{bmatrix}
z_1^y \\ z_2^x \\ z_2^y
\end{bmatrix}
\leq
\begin{bmatrix}
2\\ 1\\ 1/2\\ 3/2\\ 1/2\\ 3/2\\ 1\\ 1\\ 1\\ 1\\
\end{bmatrix}.
\end{equation}
\begin{figure}[b]
\includegraphics[width=\linewidth]{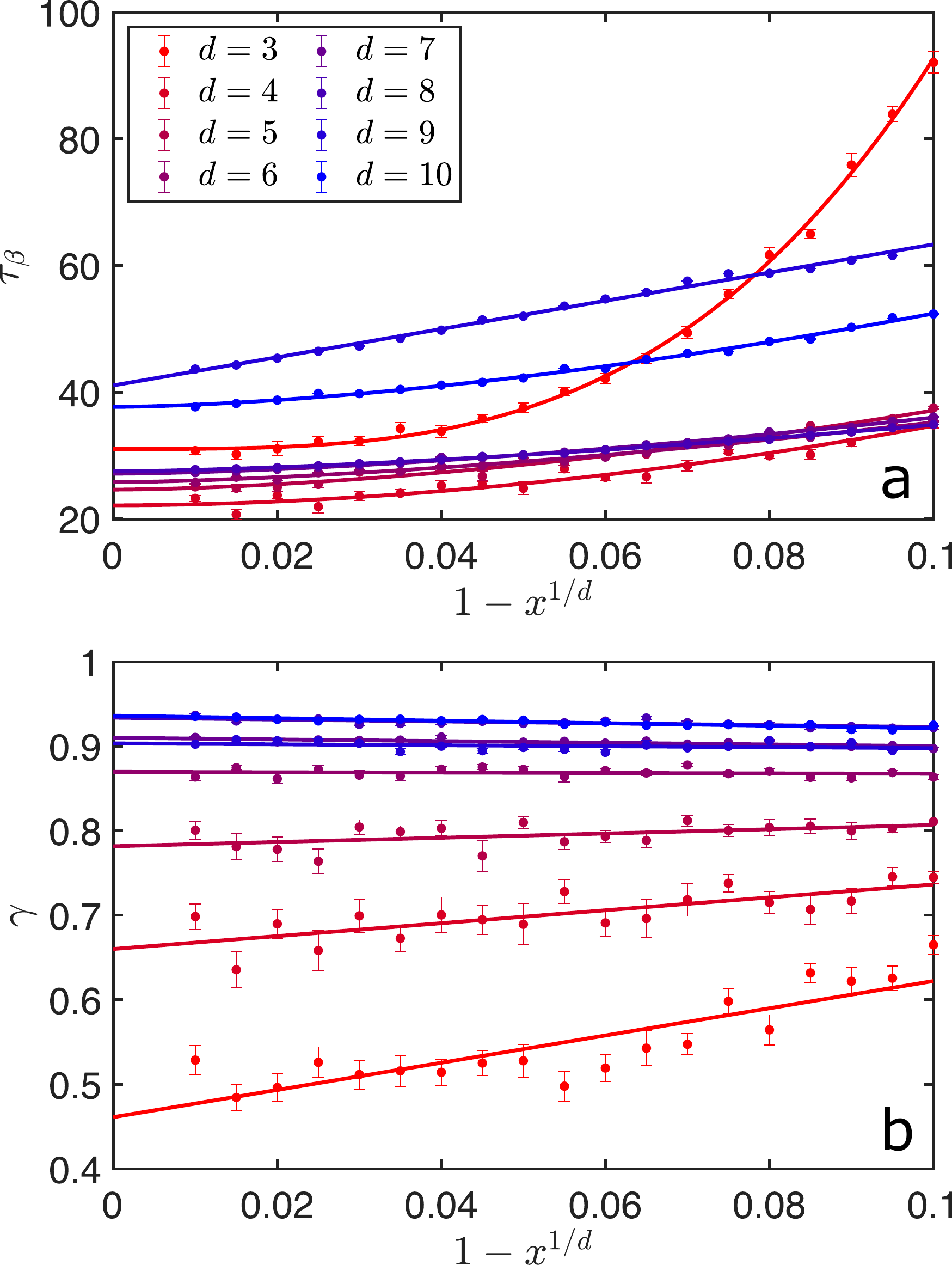}
\caption{Equilibration parameters for different $d$ and $x$. Rough empirical fitting forms are given for each as a function of the distance to crystalline close packing $1-x^{1/d}$. \textbf{a)} The relaxation time empirically scales as $\tau_\beta = c_1 + c_2(1-x^{1/d})^{\mu}$ with $1 < \mu < 4$. \textbf{b)} The stretching factor $\gamma$ scales nearly linearly, with both intercept and slope increasing with $d$ except for their marginal decrease in $d=9$.}
\label{fig:equilibration}
\end{figure}

This integral can be computed in the same way as was used for Eq.~\eqref{eq:c2ExampleInt}, but here no reflection symmetry with respect to the polynomial or the polytope symmetries exists. LattE evaluates this integral to $53/45$. Performing this same analysis on the remaining $10$ faces, and summing the results, yields $I^{(1)}_2 = 227/45$.

\begin{figure*}
\begin{minipage}[t]{.25\linewidth}
\vspace{0pt}
\centering
\begin{tabular}{c | c | c  }
\hline
\hline
$d$ & $s_c(0.98)$ & $\Xi$ \\
\hline
3 & -14.1402(13) & 13.91(11)\\
4 & -19.9812(7) & 20.6(2) \\
5 & -26.0416(11) & 27(1)\\
6 & -32.367(3) & 36(8)\\
7 & -38.878(3) & 37(17)\\
8 & -45.564(4) & 38(9)\\
9 & -50.823(2) & 50(30)\\
10 & -58.965(1) & 60(30)\\
\hline
\hline
\end{tabular}
\end{minipage}
\begin{minipage}[t]{.45\linewidth}
\vspace{0pt}
\centering
\includegraphics[width=0.9\columnwidth]{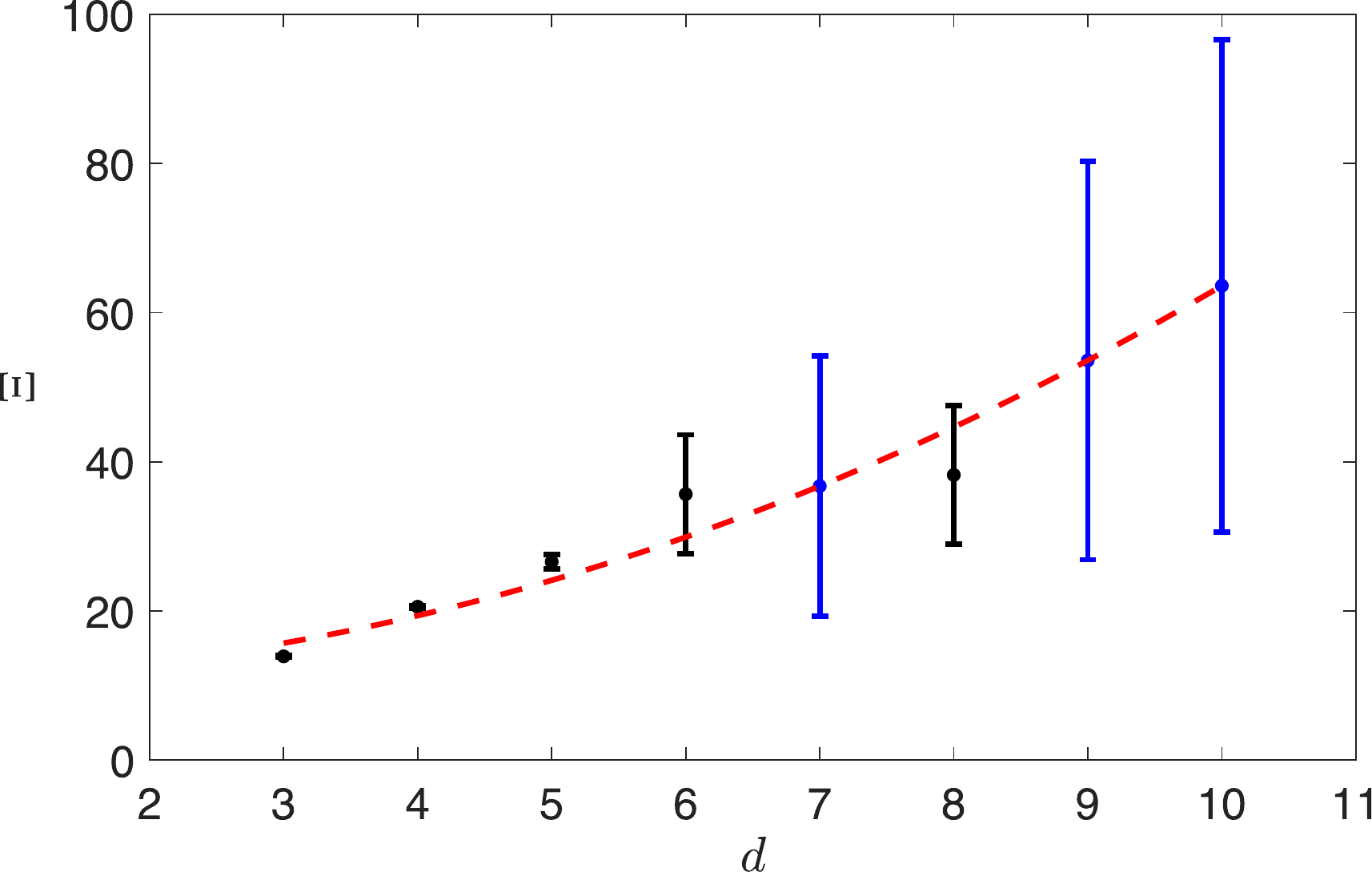}
\end{minipage}%
\caption{Entropy per particle $s_c(x_0=0.98)$  extracted from the thermodynamic extrapolation (from Eq.~\eqref{eqn:finitesize}) by Frenkel-Ladd integration for various crystals. Finite-size entropy correction coefficient from Eq.~\eqref{eqn:finitesize} at $x_0 = 0.98$ (black points). Empirically, this coefficient grows nearly quadratically  with dimension, $\Xi = 7.2(7) + 0.75(2) d^2$ (dashed line). This scaling can be used to estimate $\Xi$  (and thus $s_c$) in $d=7$ and $d=9$ and 10 (blue points), for which a single system size is computationally accessible. Error bars denote $95\%$ confidence intervals.}
\label{fig:einF}
\end{figure*}

\section{Free volume equation of state expansion}
\label{sec:freeVolumeExpansion}

In this Appendix, we provide a schematic derivation of the free volume equation of state~\cite{kirkwood_critique_1950, kamien_entropic_2007} with corrections agnostic to the crystal type and dimension, in order to show that $\kappa_0 > 0$ for all hard sphere crystals and that $\kappa_1 < 0$ is plausible for $\lambda_9$ and potentially other higher dimensional crystals.

In Eq.\ref{eq:cryst}, we wrote the approximate crystal entropy assuming that the free volume was defined by a linear cage size $a\epsilon\sigma$, where $\epsilon = 1-x^{1/d}$. This can be made more precise by instead performing an expansion of the free volume in $x$, such that
\begin{equation}
v_\mathrm{free} = \sigma^d\sum^\infty_{i=0} c_i (1-x^{1/d})^{d+i}
\end{equation}
where $c_0 = a^d > 0$. Here the sum starts from $i=0$ because it must reduce to Eq.~\eqref{eq:cryst} in the limit $x \to 1$. Note that from Fig.~\ref{fig:curvedFV}, the boundary of the free volume is concave in the limit $x \to 1$, and thus $c_1>0$.

The full expansion of the free volume allows Eq.~\ref{eq:cryst} to be rewritten as
\begin{equation}
\begin{split}
s_c =& -\ln x + \ln\bigg[ \sum_{i=0}^\infty c_i(1-x^{1/d})^{d+i}\bigg]  \\ &+\ln V_d - d \ln (\Lambda/\sigma) + 1 \ ,
\end{split}
\end{equation}
from which the equation of state follows:
\begin{equation}
\begin{split}
p =& -x\bigg(\frac{\partial s_c}{\partial x}\bigg)_\beta \\ =& 1 + \frac{\sum_{i=0}^\infty c_ix^{1/d}(1+\frac{i}{d}) (1-x^{1/d})^{d+i-1}}{\sum_{i=0}^\infty c_i(1-x^{1/d})^{d+i}} \\
=& \frac{1}{1-x^{1/d}} + \frac{c_1}{dc_0} \\ &+ \frac{2c_0c_2 - c_1(c_1 + c_0)}{dc_0^2}(1-x^{1/d}) + \dots \ .
\end{split}
\end{equation}
This yields ${\kappa_0 = \frac{c_1}{dc_0} > 0}$ and ${\kappa_1= \frac{2c_0c_2 - c_1(c_1 + c_0)}{dc_0^2}}$, which can be either positive or negative depending on the sign and magnitude of $c_2$. Hence, it is plausible that even thermodynamic $\lambda_9$ crystals could have $\kappa_1<0$.

\section{Crystal equilibration}
\label{sec:equilibrationConstants}
We empirically find that crystal MSD follow Eq.~\eqref{eqn:tauDef} with three fit parameters: the plateau height $\Delta$, which we relate to the constant $a$ in Eq.~\eqref{eq:dynamicA}; the relaxation time $\tau_\beta$, which sets the sampling time scale; and the stretching exponent $\gamma$. In Fig.~\ref{fig:equilibration}, we see that the latter two depend only weakly on $d$ and $x$, except for $d=3$. The small and decreasing $\tau_\beta$ upon approaching $\maxphi$ indicates that denser crystals relax much faster than those near coexistence. We further find that $\gamma$ is approximately linear in the distance to $\maxphi$ with both the slope and intercept increasing with dimension aside from the anomalous case of $d=9$, for which both show a marginal decrease. These observations suggest that crystal dynamics become increasingly single-particle--like as dimension increases. This dynamical observation is also consistent with the static observation that direct cell cluster expansion converges more rapidly as $d$ increases. A first-principle explanation of these features, however, is still lacking.

\section{Finite-size scaling of the reference crystal entropy}
\label{sec:finSizeRef}

From the entropy per particle $S_c/N$ for several finite system sizes, the thermodynamic entropy is obtained using a simple linear fit  (Fig.~\ref{fig:einF})
\begin{equation}
s_c = \frac{S_c}{N} - \frac{\Xi}{N}.
\label{eqn:finitesize}
\end{equation}
Note that for $d=7$, $9$, and $10$ a single crystal size is computationally available ($N=17496$, $39366$, and $81920$ respectively). Because the proportionality constant $\Xi$ scales roughly quadratically with $d$ for fixed $x$, we nevertheless interpolate its value for $d=7$, and extrapolate it for $d=9$ and $10$, in order to estimate $s_c$ in these dimensions as well. In $d=9$, however, where soft modes give rise to additional finite-size corrections, this extrapolation is particularly unreliable.

\bibliography{liquidCrystalCoexistence,footnotes}
\end{document}